\documentclass[useAMS,usenatbib, fleqn]{mn2e}
\usepackage{graphicx}
\usepackage{amsmath,amssymb}
\title[Conditions for HD Cooling in the First Galaxies Revisited]{Conditions for HD Cooling in the First Galaxies Revisited: 
Interplay between Far-Ultraviolet and Cosmic Ray Feedback in Population III Star Formation}
\author[Nakauchi et al.]{Daisuke Nakauchi$^{1}$
, Kohei Inayoshi$^{1}$, and Kazuyuki Omukai$^{1, 2}$\\
$^1$Department of Physics, Kyoto University, Sakyo, Kyoto 606-8502, Japan\\
$^2$Astronomical Institute, Tohoku University, Aoba, Sendai 980-8578, Japan
}

\begin{document}

\maketitle


\begin{abstract}
HD dominates the cooling of primordial clouds with enhanced ionization, e.g. shock-heated clouds in structure formation or supernova remnants, relic HII regions of Pop III stars, and clouds with cosmic-ray (CR) irradiation.
There, the temperature decreases to several 10 K and the characteristic stellar mass decreases to $\sim 10\ {\rm M}_{\odot}$, in contrast with first stars formed from undisturbed pristine clouds~($\sim 100\ {\rm M}_{\odot}$).
However, without CR irradiation, even weak far ultra-violet (FUV) irradiation suppresses HD formation/cooling.
Here, we examine conditions for HD cooling in primordial clouds including both FUV and CR feedback.
At the beginning of collapse, the shock-compressed gas cools with its density increasing, while the relic HII region gas cools at a constant density.
Moreover, shocks tend to occur in denser environments than HII regions.
Owing to the higher column density and the more effective shielding, the critical FUV intensity for HD cooling in a shock-compressed gas becomes $\sim 10$ times higher than in relic HII regions.
Consequently, in the shock-compressed gas, the critical FUV intensity exceeds the background level for most of the redshift we consider~($6 \la z \la 15$), while in relic HII regions, HD cooling becomes effective after the CR intensity increases enough at $z \la 10$.
Our result suggests that less massive ($\sim 10\ {\rm M}_{\odot}$) Pop III stars may be more 
common than previously considered and could be the dominant population of Pop III stars.
\end{abstract}


\section{Introduction}\label{sec:intro}
The first generation of stars are believed to have had a significant impact on the cosmic 
evolution through their radiative, mechanical and chemical feedback~\citep[e.g.,][]{Ciardi2005}.
Stellar extreme-ultraviolet (EUV) photons~($> 13.6$ eV) ionize the circumstellar gas 
to form an HII region, thereby initiating the process of the reionization of the intergalactic medium~(IGM).
Even the single supernova (SN) of a massive ($\ga 150{\rm M}_{\odot}$) first-star is able to totally 
evacuate the gas and suppresses subsequent star formation in the same minihalo~\citep{Bromm2003,Wada2003} 
although some of the materials blown out by the SN of a less-massive 
($20\mbox{-}40{\rm M}_{\odot}$) progenitor may fall back again and 
commence another episode of star formation~\citep{Ritter2012}.
The ejected gas by a SN contributes to the metal enrichment of the IGM.
The extent of such feedback depends crucially on the initial mass function (IMF), as well as 
the star formation rate (SFR), of first stars.

Stars formed from the metal-free primordial gas, or Population III (Pop III) stars, 
have been classified into two sub-populations depending on 
whether the gas is still in an undisturbed pristine state
with a low electron fraction $\sim 10^{-4}$~(Pop III.1), 
or the gas has already been affected by the radiative or kinematic feedback 
from pre-existing stars~(Pop III.2), following the nomenclature proposed by O'Shea et al. (2008)\footnote{Previously, Pop III.2 stars were called in different names from authors to authors, 
for example, Pop II.5 stars (Mackey, Bromm \& Hernquist 2003).}.
Despite with the same primordial composition, the thermal evolution of the gas
is fairly different in those cases, resulting in different characteristic stellar masses.
In the standard $\Lambda$CDM cosmology,  
Pop III.1 stars are predicted to be formed in minihalos of $\sim 10^6\ {\rm M}_{\odot}$
~\citep{Couchman1986,Haiman1996b,Tegmark1997}.
During its protostellar collapse, cooling is entirely by ${\rm H}_2$, which is 
produced by the following electron-catalyzed reaction~\citep[so-called the ${\rm H}^-$ channel;][]{Peebles1968, Hirasawa1969}:
\begin{equation}
{\rm H} + e \rightarrow {\rm H}^- + \gamma, 
\label{eq:H-1}
\end{equation}
\begin{equation}
{\rm H}^- + {\rm H} \rightarrow {\rm H}_2 + e. 
\label{eq:H-2}
\end{equation}
With the high value of the minimum temperature $200\ \rm K$ attained at the density $\sim 10^{4}\ {\rm cm^{-3}}$, 
the fragmentation mass is as large as $\sim 10^{3}\ {\rm M_{\odot}}$
~\citep{Bromm1999, Bromm2002, Abel2002, Yoshida2006}, 
which is set by the Jeans mass at this epoch.
In this case, the final mass of formed stars is set by the radiative feedback from the growing protostar, 
and its value is a few 10 - a few 100 ${\rm M_{\odot}}$, much higher than the present-day counterpart~\citep{Omukai_Palla2001,Omukai_Palla2003,McKee2008,Yoshida2008, Hosokawa2011, Stacy2012, Susa2013, Hirano2014}.
On the other hand, Pop III.2 stars are formed from the metal-free gas affected either by 
the radiative or kinematic feedback from earlier generation of stars, typically in atomic-cooling halos 
with virial temperatures $\gtrsim 10^4\ {\rm K}$~\citep{O'Shea2008b,Greif2008}.
An example of Pop III.2 stars is those formed in the relic HII regions of defunct Pop III 
stars~\citep{Kitayama2004, Whalen2004, Yoshida2007a, Abel2007}.
In this case, a larger amount of ${\rm H}_2$ is produced~($\sim 10^{-3}$) via the ${\rm H}^-$ channel~(Eqs. \ref{eq:H-1} and \ref{eq:H-2}) than in the Pop III.1 case~($\sim 10^{-4}$) \citep[e.g.,][]{O'Shea2005}.
Once the temperature plummets below $150\ \rm K$ by this ${\rm H}_2$ cooling, deuterium is rapidly converted into HD via the exothermic reaction,
\begin{equation}
{\rm D}^+ + {{\rm H}_2} \rightarrow {\rm HD} + {\rm H}^+.
\label{eq:HD}
\end{equation}
This cooling by HD lowers the temperature further to $\sim 30\ \rm K$~\citep[e.g.,][]{Nagakura2005}.
The characteristic mass of stars formed in such an environment is predicted to be $\sim 10\ {\rm M_{\odot}}$~\citep{Uehara2000, Nakamura2002,Mackey2003,Machida2005,Nagakura2005,Johnson2006,Yoshida2007a, Yoshida2007b,Wolcott2011,Hosokawa2012}.
That is, Pop III.2 stars may be typically less massive 
than preceding Pop III.1 stars although they are more massive than ordinary present-day stars
\footnote{It should be noted that recent numerical simulations suggest that there is another possible pathway of lower-mass primordial star formation in Pop III.1 case, like disk fragmentation~\citep[e.g.,][]{Greif2012}. So the typical mass-scale of Pop III.1 stars remains an open question.}.

With high effective temperatures ($\sim 10^5\ {\rm K}$)~\citep{Schaerer2002}, massive Pop III 
stars emit a copious amount of far-UV (FUV) photons in the Lyman-Werner (LW) bands ($11.2 \mbox{-}13.6$ eV).
The FUV radiation field photodissociates ${\rm H}_2$, thereby prohibiting efficient cooling in the primordial gas.
Even a single massive Pop III star emits enough FUV photons to suppress the subsequent star formation 
in the same halo~\citep{Omukai1999,Glover2001}.
In addition, the gas in a minihalo is fairly vulnerable to the external irradiation of FUV photons and 
thus the number of Pop III.1 stars formed can be severely regulated by the FUV background~\citep[e.g.,][]{Haiman1997}.
On the other hand, Pop III.2 star formation in more massive halos still goes on by atomic cooling 
even under the FUV background and is not strongly affected.~\citep[Haiman et al. 1997;][]{O'Shea2008b}.
This indicates the possibility that Pop III.2 stars can be the major population of primordial stars and 
significant sources of radiative and chemical feedback to subsequent star formation~\citep{Trenti2009,Souza2011}.
Moreover, the Pop III.2 star-formation epoch may have lasted until redshifts as low as 
$z \sim 6$~\citep{Tornatore2007,Souza2011,Johnson2013}.
If so, gamma-ray bursts (GRBs) or SNe of Pop III.2 stars are detectable with future facilities, 
which allow us to peer into the nature of primordial stars~\citep{Nakauchi2012, Whalen2013, Tanaka2013}.

Later, however, it was found that HD formation is strongly suppressed in the presence of 
even a weak FUV radiation field~\citep[Yoshida et al. 2007b;][]{Wolcott2011} 
since the slight photodissociation of ${\rm H}_2$ causes the significant reduction of the HD fraction.
Without HD cooling, the thermal evolution of a Pop III.2 star-forming gas becomes identical 
to the Pop III.1 case, resulting in Pop III.2 stars having similar masses to Pop III.1 stars.
However, this is not yet the conclusive result since the previous studies have not properly taken account of 
the following effects, which favors HD formation/cooling.

In the galaxy formation epoch, shocks occur ubiquitously, for example, associating 
with the virialization of halos or SN explosions. 
In a shocked gas, the enhanced ionization degree enables efficient H$_2$/HD formation and cooling
~\citep{Uehara2000,Mackey2003,Johnson2006,Greif2008}.
For a strong shock, the post-shock gas is compressed isobarically with its cooling, 
i.e., the temperature decreases with increasing the density ~\citep[e.g.,][]{Shapiro1987}.
On the other hand, for the free-falling cloud in a relic HII region,  
rapid cooling makes the temperature to plummet almost isochorically, 
i.e., the temperature decreases at a constant density~\citep[Yoshida et al. 2007b;][]{Johnson2006, Wolcott2011}.
With the higher column density, the FUV radiation field is more effectively shielded for the post-shock flow than in the relic HII region, resulting in the lower FUV flux in the former case. 
Nonetheless, in the previous studies~\citep[e.g.,][]{Wolcott2011}, the post-shock flows are treated
without the proper account of the isobaric nature of evolution.

Associated with star-formation activities, not only the FUV background, but also the
cosmic ray~(CR) background is expected to be generated by way of 
the CR acceleration in SN remnants.
CR irradiation promotes ${\rm H}_2$ and HD formation/cooling by 
enhancing the ionization degree~\citep{Jasche2007,Stacy2007, Inayoshi2011}.
Thus, CR irradiation is able to counteract with the negative feedback of FUV radiation. 
However, only either FUV or CR irradiation has been considered in the previous studies
~\citep{Jasche2007,Stacy2007,Wolcott2011} and their combined effects have not been fully understood.

In this paper, we study the conditions for HD cooling to become important in Pop III.2 star formation 
under the presence of both FUV and CR backgrounds. 
We calculate the temperature and chemical evolution of (i) a cloud compressed by plane-parallel steady 
shocks~\citep{Shapiro1987}, and of (ii) a free-falling cloud in a relic HII region.
We find that the critical value for FUV intensity above which HD cooling is quenched 
is about an order of magnitude higher in the former than in the latter case.
Also the critical FUV intensity increases with increasing CR intensity.
We also estimate the background FUV and CR intensities from theoretically predicted values for the SFR 
in the high-redshift universe. 
By comparing the background FUV intensity with the critical value, 
we conclude that the HD mode of Pop III.2 star formation is fairly common at the galaxy formation epoch. 

The rest of this paper is organized as follows.
In \S \ref{sec:model}, we describe the numerical model and initial conditions for our calculation.
In \S \ref{sec:result}, we present the results for the free-falling clouds in relic HII regions 
and for the clouds compressed either by the virialization shock or by the SN blast wave. 
We also present the critical FUV intensity for HD cooling as a function of the CR intensity.
In \S \ref{sec:estimate}, we estimate the background FUV and CR intensities in high-redshift universe, 
and discuss whether the HD cooling condition is satisfied at these epochs.
After discussing the implications and uncertainties of our study in \S \ref{sec:discussion}, 
we summarize our findings in \S \ref{sec:conclusion}.

\section{Model}\label{sec:model}
In this section, we present our model for calculating the thermal and chemical evolution of primordial clouds.

\subsection{Formulation}
Here, we first describe the method of calculation for the free-falling cloud in a relic HII region 
and then for the shock-compressed gas in structure formation or an SN explosion.
We use the one-zone model, where we focus on the evolution of the central core for the free-falling cloud and of the coolest gas layer for the shock-compressed gas before the self-gravity becomes effective by using the plane-parallel and steady shock approximation.

\subsubsection{Free-falling Cloud in a relic HII region}\label{subsec:free-fall_form}
When a Pop III star dies and collapses directly to a black hole (BH), the gas in the relic HII region surrounding the star begins to recombine and cool to form a subsequent star~\citep{Kitayama2004, Whalen2004, Yoshida2007a, Yoshida2007b, Abel2007}. Here, we assume that the formed BH has no influence on the evolution of the gas in the relic HII region for simplicity.
At first, the baryon mass in the relic HII region is too small for the gas to contract by self-gravity, but the gas loses pressure-support by efficient radiative cooling and contracts being attracted by the gravitational potential of the DM halo. 
After the temperature decreases so that the Jeans mass becomes lower than the HII region mass, the gas begins to contract in the almost free-fall way by self-gravity.
Owing to its high pressure, the gas in the HII region is accelerated monotonically with radius up to a supersonic velocity and expands into the interstellar medium~(Whalen et al. 2004; Kitayama et al. 2004).
However, our model does not take into account the initial velocity profile within the HII region for simplicity.
The following prescription is the same as the previous studies~(Johnson \& Bromm 2006; Yoshida, Omukai \& Hernquist 2007; Wolcott-Green \& Haiman 2011).

An isolated cloud collapsing at the rate close to free-fall by its self-gravity 
develops a core-envelope structure~\citep{Larson1969, Penston1969}, consisting of a nearly Jeans-scale core with the
constant density and an envelope where the density decreases radially as $\rho \propto r^{-2}$.
In our one-zone model, we follow the evolution of the physical quantities at the core center~\citep{Omukai2001}.
Since collapse proceeds roughly at the free-fall rate, the density evolves as
\begin{equation}
\frac{d \rho}{dt} = \frac{\rho}{t_{\rm ff}},   
\label{eq:drho_dt}
\end{equation}
where $t_{\rm ff} = \sqrt{{3 \pi}/{32 G \rho}}$ is the free-fall time and $G$ the gravitational constant.
Temperature evolution follows the energy equation:
\begin{equation}
\frac{d e}{dt} = - P \frac{d}{dt}\left(\frac{1}{\rho}\right) - \frac{\Lambda_{\rm net}}{\rho},   
\label{eq:energy}
\end{equation}
where $e$ is the internal energy per unit mass, $P$ the pressure,  
$\Lambda_{\rm net}$ the net cooling rate per unit volume.
To supplement the equations above, 
the ideal-gas equation of state $P = \rho k_{\rm B} T / \mu m_{\rm H}$ is used, 
where $k_{\rm B}$ is the Boltzmann constant, $m_{\rm H}$ the proton mass and $\mu$ the mean molecular weight.
Using the ratio of specific heat $\gamma$, specific energy is related to pressure as 
$e = P/\rho (\gamma -1)$.
Throughout this paper, we take $\mu = 1.2$ and $\gamma = 5/3$ neglecting molecular contribution.
Considering that the size of the contracting core is half the Jeans length $\lambda_{\rm J}=\sqrt{{\pi k_{\rm B} T}/{G \rho \mu m_{\rm H}}}$, we calculate the column density of the core as $N_{\rm J} = n_{\rm H} \lambda_{\rm J} / 2$~\citep{Inayoshi2011}.
In this case, the initial size of this region is estimated as $\sim 1$ kpc, which is an overestimation compared to the typical size of an HII region $R_{\rm HII} \sim 100$ pc~(Whalen et al. 2004; Kitayama et al. 2004).
However, at the beginning of collapse, owing to the low gas density and H$_2$ fraction, the H$_2$ column density will be too low for effectively shielding the gas from the FUV filed, and the above overestimation will not matter for estimating the shielding factor.

In our model of free-fall collapse, the collapse equation and the energy equation (Eqs. \ref{eq:drho_dt}, \ref{eq:energy}) are decoupled and feedback from thermal pressure is not included for simplicity.
In the runaway-collapse phase, an isothermal cloud collapses approximately at a free-fall rate~\citep{Larson1969, Penston1969}.
Although the pressure gradient is not negligible, the deviation of the collapse time from free fall is only by a factor of 1.58~\citep{Larson1969}.
For a non-isothermal case, the deviation is represented as a monotonically increasing function of the effective adiabatic index $\gamma$ of the cloud in equations 7-9 of \cite{Omukai2005}.
For example, the collapse time becomes $\sim 3$ times longer than the free-fall time for $\gamma = 1.2$.
If we use this generalized collapse time for the collapse equation, the collapse equation 4 is coupled with the energy equation 5.
However, we find that the effective adiabatic index becomes smaller than 1.2 for most of the evolutionary trajectory, and the deviation from free-fall may be at most by a factor of a few.
Thus, feedback from thermal pressure may change our results little.

Here, we consider that an overdense region collapsing within an HII region is large enough. Then, since the overdense region is strongly bounded by the gravitational field of DM and gas, it contracts by the gravitational force which exceeds the pressure force from its less-dense surroundings.
Therefore, the compression of the overdense region by external pressure may be neglected, and it will collapse roughly at a free-fall rate.

Although our formulation does not include DM gravity, it becomes significant at the beginning of gas evolution. Then, we recalculate the evolution of the gas in relic HII regions taking account of DM gravity by modifying $\rho$ in $t_{\rm ff}$ of Eq. \eqref{eq:drho_dt} to $\rho + \rho_{\rm DM}$, but we find little change in our results.
Here, $\rho_{\rm DM} = 1.3 \times 10^{-24}\ {\rm g}\ {\rm cm}^{-3}$ is the mean DM mass density of a virialized halo at $z=15$.

\subsubsection{Shock-Compressed Gas}\label{subsec:shock_form}
After the passage of a blast wave, which is generated in structure formation or an SN explosion, the post-shock gas has a very high temperature due to shock heating and it cools and contracts with the radiative cooling time. At first, self-gravity is not effective in the shocked gas layer. We focus on the evolution of the most cooled gas layer, assuming that the post-shock gas layer is plane-parallel and the flow is steady~\citep{Shapiro1987, Yamada1998, Inayoshi2012}.
The initial condition of a shock-compressed gas is characterized by two parameters: the pre-shock density $\rho_0$ and 
the velocity of the shock front $v_0$.
In the strong shock limit, the physical quantities just behind the shock front are given by the Rankine-Hugoniot relation:
\begin{equation}
\rho_1 = 4 \rho_0,
\label{eq:rho1}
\end{equation}
\begin{equation}
v_1 = \frac{1}{4} v_0, 
\label{eq:v1}
\end{equation}
\begin{equation}
T_1 = \frac{3}{16}\frac{\mu m_{\rm H}}{k_{\rm B}}v_0^2
\end{equation}
\begin{equation}
\sim 1.1 \times 10^4 \left(\frac{\mu}{1.2}\right) \left(\frac{v_0}{20\ {\rm km}/{\rm s}}\right)^2 {\rm K}, \notag
\label{eq:T1}
\end{equation}
where the subscripts 0 and 1 refer to the pre- and post-shock quantities, respectively.
Behind the shock front, the flow evolution is described by 
\begin{equation}
\rho_1 v_1 = \rho v,
\label{eq:rho_v}
\end{equation}
\begin{equation}
\rho_1 v_1^2 + P_1 = \rho v^2 + P,
\label{eq:rho_v^2}
\end{equation}
along with the energy equation (\ref{eq:energy}).
The column density of the post-shock gas can be estimated from
\begin{equation}
N_{\rm H} (t) = n_{{\rm H}, 0} v_0 t,
\label{eq:clden_b}
\end{equation}
where $n_{{\rm H}, 0}$ is the number density in the pre-shock gas and $t$ is the time since the shock occurs.

Rapid cooling in the post-shock layer leads to the formation of a dense gas shell, which eventually becomes 
gravitationally unstable and fragments. 
For the SN shock, we assume that fragmentation occurs when the column density of the post-shock layer $N_{\rm H}$ 
exceeds that in the Jeans length $N_{\rm J}$ \citep{Safranek-Shrader2010, Inayoshi2012}:
\begin{equation}
N_{\rm H} (t) = N_{\rm J}.
\label{eq:cond_sn}
\end{equation}
For the structure-formation shock, the mass of the post-shock layer is doubled compared to the SN case, since in one dimension, two flows collide to form the shocked region and two shock fronts propagate in opposite directions to each other. In this case, fragmentation occurs under the condition:
\begin{equation}
2N_{\rm H} (t) = N_{\rm J}.
\label{eq:cond_sf}
\end{equation}
In addition to them, for fragmentation, the growth time of a perturbation, which is approximately given by 
the local free-fall time $t_{\rm ff}$, must be shorter than the contraction time of the layer, 
i.e., the cooling time $t_{\rm cool}$ (Yamada \& Nishi 1998). 
We have confirmed this condition is always satisfied at the fragmentation epoch determined by 
the column-density conditions given above.
After fragmentation, each fragment or clump begins to contract by its self-gravity.
We also follow its evolution by the same one-zone model as in Sec. \ref{subsec:free-fall_form}.

\subsection{Thermal and Chemical Processes}\label{subsec:thermal_chemical}
The net cooling rate $\Lambda_{\rm net}$ in Eq. \eqref{eq:energy} consists of 
the following terms: 
\begin{equation}
\Lambda_{\rm net} = \Lambda_{\rm H} + \Lambda_{{\rm H}_2} + \Lambda_{\rm HD} - \Gamma_{\rm CR}, 
\label{eq:lambda_net}
\end{equation}
where $\Lambda_{\rm H}$, $\Lambda_{{\rm H}_2}$ and $\Lambda_{\rm HD}$ are 
the radiative cooling rates by H Lyman-$\alpha$ line emission, ${\rm H}_2$-line emission and HD-line emission, 
respectively, and $\Gamma_{\rm CR}$ is the CR heating rate.
We adopt the cooling rates $\Lambda_{\rm H}$ from \cite{Glover2007}, $\Lambda_{{\rm H}_2}$ from \cite{Galli1998}, and $\Lambda_{\rm HD}$ from \cite{Galli2002}. 
Since each CR ionization associates with heating of $\sim 6\ {\rm eV}$ \citep{Spitzer1969}, 
\begin{equation}
\Gamma_{\rm CR} \sim 9.6 \times 10^{-12}\ n({\rm H})\ \zeta_{\rm CR}\ {\rm erg}\ {\rm cm}^{-3}\ {\rm s}^{-1},
\label{eq:cr_heat}
\end{equation}
where $\zeta_{\rm CR}$ is the hydrogen ionization rate by CRs, 
and $n({\rm H})$ is the number density of hydrogen atoms.
At densities we consider~($\lesssim 10^{7}\ {\rm cm}^{-3}$), the contribution from 
chemical cooling and heating is small and is neglected in this paper~\citep{Omukai2001}.

Our chemical network includes following 11 species: H, ${\rm H}_2$, $e$, ${\rm H}^+$, ${\rm H}_2^+$, ${\rm H}^-$, D, HD, ${\rm D}^+$, ${\rm HD}^+$ and ${\rm D}^-$.
The deuterium fraction is set to $4.0 \times 10^{-5}$, and the He chemistry is neglected since it is thermally inert 
in the temperature range we consider.
Chemical reactions include 18 hydrogen collisional reactions in Table 1 of \cite{Yoshida2006}, 
18 deuterium collisional reactions in Table 1 of  \cite{Nakamura2002},
photo-dissociation of H$_2$, HD, and H$^{-}$, and CR ionization.
For collisional reactions, we use the updated rate coefficients by \cite{Glover2008}.
The photo-dissociation rate coefficients are taken from \cite{Omukai2001} and 
\cite{Wolcott2011}, and the ${\rm H}_2$ and HD shielding factors are from~\citet{Wolcott2011}.
For the shielding factors, we take account of the Doppler shift due to the bulk motion of the cloud 
as in \cite{Omukai2001} for the free-falling case and in \cite{Omukai2007} for the shock-compressed case.
The stellar radiation field is assumed to have the diluted black-body spectrum with the temperature of $10^4$\ K, resembling that from a Pop II star cluster, and its intensity is parameterized by $J_{21} \equiv J_{\rm \nu_{Ly}}/10^{-21}\ {\rm erg}\ {\rm cm}^{-2}\ {\rm s}^{-1}\ {\rm Hz}^{-1}$, where $J_{\rm \nu_{Ly}}$ is the intensity at the Lyman limit.
The CR ionization rate of neutral hydrogen $\zeta_{\rm CR}$ is treated as a free parameter, 
since the CR intensity at the high-redshift universe is highly uncertain.
However, the CR intensity may be related with the star-forming activity, and the SFR at $z \sim 10$ is theoretically 
estimated to be no more than the present-day value~(e.g., Tornatore et al. 2007; Trenti \& Stiavelli 2009; Johnson et al. 2013).
We thus consider the CR intensity in the range of $\zeta_{\rm CR}\lesssim 10^{-17}\ {\rm s}^{-1}$, which are smaller than the Galactic value $10^{-17}\ {\rm s}^{-1} \lesssim \zeta_{\rm CR} \lesssim 10^{-15}\ {\rm s}^{-1}$~\citep{Hayakawa1961,Spitzer1968,Webber1998,McCall2003,Indriolo2007}.

\subsection{Initial Conditions}
\subsubsection{Relic HII Region}\label{subsec:free-fall_IC}
In the relic HII region of a defunct Pop III star, a primordial cloud begins to collapse from a highly ionized state.
The temperature and density in the relic HII region are almost uniform with $1\mbox{-}3 \times 10^4\ {\rm K}$ and 
$0.1\mbox{-}1\ {\rm cm}^{-3}$~(Kitayama et al. 2004; Whalen et al. 2004; Abel, Wise \& Bryan 2007; Yoshida et al. 2007a).
We thus adopt $T_0 = 3 \times 10^4\ {\rm K}$ and $n_{\rm H, 0} = 0.3\ {\rm cm}^{-3}$ as the fiducial initial values, 
and also study the cases with higher and lower densities $n_{\rm H, 0} = 0.03, 3 \ {\rm cm}^{-3}$ to see 
the dependence on the initial density.
We set the initial chemical abundances as $y_0({\rm H}^{+}) = 1.0$, $y_0({\rm D}^{+}) = 4.0 \times 10^{-5}$, $y_0(e) = y_0({\rm H}^{+}) + y_0({\rm D}^{+})$, and $y(i) = 0$ for other species, where 
$y(i)$ is the fractional abundance of the $i$-th species to the hydrogen nuclei.

\subsubsection{Shock-Compressed Gas}\label{subsec:shock_IC}
We consider two kinds of shocks which occur in the virialization of halos and SN explosions.
Here, we need to specify the shock velocity $v_0$, pre-shock density $n_{{\rm H}, 0}$, 
and initial chemical abundance $y_0(i)$ for each case.

For the virialization shock, 
the mean gas number density within the virialized halo is estimated as \citep{Clarke2003}
\begin{equation}
n_{{\rm H}, 0} \sim 0.13 \left(\frac{1+z}{15}\right)^{3} {\rm cm}^{-3}.
\label{eq:n_H_halo}
\end{equation}
However, according to the recent studies on first galaxy formation, 
the shock front does not sustain at the virial radius and its location shrinks inward owing to 
efficient Ly-$\alpha$ emission~\citep{Wise2007,Wise2008}.
Protogalaxies accrete gas via cold flows in dense filaments, with the infall velocity of $\sim 20\ {\rm km}\ {\rm s}^{-1}$.
Such cold flows penetrate deep inside the halo and collide each other to form virialization shocks 
with pre-shock densities $1\mbox{-}10^3\ {\rm cm}^{-3}$, much higher than in Eq. \eqref{eq:n_H_halo}.
Considering this effect, we adopt $(v_0, n_{\rm H, 0}) = (20, 1.0)$ as the fiducial values for the pre-shock gas.

The shocked gas behind a SN blast wave cools and is compressed significantly after the SN remnant enters 
the pressure-driven expansion phase \citep{Machida2005, Nagakura2009, Chiaki2013}.
The shock velocity early in this phase is $\sim 100\ {\rm km}\ {\rm s}^{-1}$~(Machida et al. 2005; Nagakura et al. 2009; Chiaki et al. 2013). 
The pre-shock density is comparable to the density in relic HII regions $0.1\mbox{-}1\ {\rm cm}^{-3}$, since core-collapse SNe explode within HII regions which their progenitors produce.
Hence, we adopt $(v_0, n_{\rm H, 0}) = (100, 0.1)$ as the fiducial values for the SN shock.

\cite{Shapiro1987} and \cite{Kang1992} studied the chemical composition in the primordial gas 
just behind a shock front. We adopt their results as the initial composition in the shock-compressed gas,  
which are summarized in Table \ref{tab:shock_comp} for the virialization and SN shocks.
The fractions of species not presented in Table \ref{tab:shock_comp} are set to zero, 
except for the electron fraction $y_0(e) = y_0({\rm H}^{+}) + y_0({\rm D}^{+})$. 

\begin{table}
\caption{Initial chemical composition in the shock-compressed gas}
\begin{center}
\begin{tabular}{c|cccc}
\hline
$v_0$\ (km\ ${\rm s}^{-1}$) &   $y_0({\rm H})$         &  $y_0({\rm H}^{+})$ & $y_0({\rm D})$  & $y_0({\rm D}^{+})$    \\ \hline
20                                        &  0.99     &  0.01                & $3.96 \times 10^{-5}$ & $4.0 \times 10^{-7}$       \\ 
100                                      &  0.3       &  0.7                  & $1.2 \times 10^{-5}$ & $2.8 \times 10^{-5}$        \\
\hline
\end{tabular} 
\end{center}
\label{tab:shock_comp}
\end{table}

\section{Results}\label{sec:result}
\subsection{Relic HII region}\label{subsec:free-fall_result}
Here, we present the results for the free-falling cloud in a relic HII region.
We also find the critical FUV intensity above which HD cooling is suppressed 
for the different values of the CR intensity.

\subsubsection{Thermal Evolution}\label{subsec:free-fall_evolve}
\begin{figure}
\begin{center}
\includegraphics[scale=1.0]{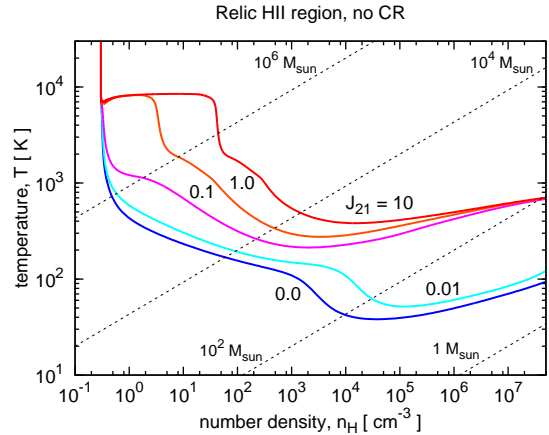}
\caption{FUV effect on thermal evolution of clouds in relic HII regions.
No CR irradiation is considered. 
The initial density is $n_{\rm H, 0} = 0.3\ {\rm cm}^{-3}$.
Different lines correspond to the different FUV intensities: 
$J_{21}=0.0, 0.01, 0.1, 1.0$, and $10$ from bottom to top.
The thin dashed lines indicate the constant Jeans masses.}
\label{fig:h2_nT_CR0}
\end{center}
\end{figure}

\begin{figure}
\begin{center}
\includegraphics[scale=1.0]{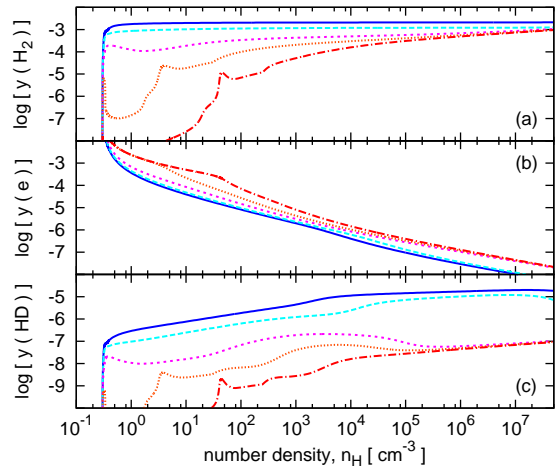}
\caption{Chemical fractions of (a) ${\rm H}_2$, (b) $e$, and (c) HD in the same clouds 
as in Fig. \ref{fig:h2_nT_CR0}.
In each panel, different lines correspond to the different FUV intensities: $J_{21}=0.0$ (solid), $0.01$ (long-dashed), $0.1$ (short-dashed), $1.0$ (dotted), and $10$ (dot-dashed).}
\label{fig:h2_nT_CR0_chem}
\end{center}
\end{figure}

First, we see the results without CR irradiation to extract the FUV effect alone.
Fig. \ref{fig:h2_nT_CR0} shows the temperature evolution, and Fig. \ref{fig:h2_nT_CR0_chem} shows the fractions of (a) ${\rm H}_2$, (b) electron, and (c) HD, for five different FUV intensities: $J_{21}=0, 0.01, 0.1, 1$, and $10$, both as a function of 
the number density $n_{\rm H}$.
The temperature first drops isochorically to 
$\sim 8000\ {\rm K}$ via efficient Ly-$\alpha$ cooling in all the cases in Fig. \ref{fig:h2_nT_CR0}, and the subsequent behavior depends on the FUV intensities.
The same results are obtained in Johnson \& Bromm (2006), Yoshida, Omukai \& Hernquist (2007), and Wolcott-Green \& Haiman (2011).

In the absence of FUV irradiation ($J_{21} = 0$), abundant ${\rm H}_2$ 
with the fraction $\sim 10^{-3}$~forms via the H$^-$ channel (Eqs. \ref{eq:H-1} and \ref{eq:H-2}) 
owing to the high initial ionization degree~(solid line in Fig. \ref{fig:h2_nT_CR0_chem}a).
The cooling by this enhanced H$_2$ makes the temperature to fall below $\simeq 200\ {\rm K}$, 
the minimum value attainable in the primordial pristine gas, where the ionization degree is 
only $\sim 10^{-4}$~\citep{Omukai2001}.
Once the temperature reaches $\la 150\ {\rm K}$, 
the exothermic reaction (Eq. \ref{eq:HD}) converts
most of the deuterium into HD (solid line in Fig. \ref{fig:h2_nT_CR0_chem}c). 
HD cooling further drops the temperature to 
the minimum value $\sim 30\ {\rm K}$ at density $\sim 10^{5}\ {\rm cm}^{-3}$, 
which is the critical density for HD to reach local thermodynamic equilibrium. 
Toward higher density, the temperature increases gradually by compressional heating.
At the temperature minimum, the parent cloud is considered to fragment into 
clumps with the Jeans mass of $\la 100\ {\rm M}_{\odot}$~\citep[Yoshida et al. 2007b;][]{McGreer2008}.
We call hereafter this mode of star formation as {\it HD-Pop III star formation}, in contrast to 
{\it H$_2$-Pop III star formation}, where H$_2$ is the sole coolant during proto-stellar collapse.

With FUV irradiation, the temperature becomes higher owing to the photodissociation of the coolants, 
H$_2$ and HD~(Fig. \ref{fig:h2_nT_CR0_chem}a, c). 
With $J_{21}$ as low as 0.01, the evolutionary track is similar to that of $J_{21} = 0$, 
although with slightly higher temperature.
With intensity $J_{21} = 0.1$, the ${\rm H}_2$ fraction is reduced to $\sim 10^{-4}$, 
and the temperature does not reach the regime of efficient HD formation/cooling 
(short-dashed line in Fig. \ref{fig:h2_nT_CR0_chem}a, c).
With the even higher FUV intensity $J_{21} \geq 1.0$, 
photodissociation strongly reduces the H$_2$ fraction at low densities 
(dotted and dot-dashed lines in Fig. \ref{fig:h2_nT_CR0_chem}a).
The clouds contract isothermally at $\sim 8000\ {\rm K}$ 
by Ly-$\alpha$ cooling until $n_{\rm H} \sim 10\ {\rm cm^{-3}}$ for $J_{21}= 1.0$ 
($\sim 100\ {\rm cm^{-3}}$ for $J_{21} =10$, respectively), where the H$_2$ self-shielding 
against FUV radiation becomes effective.
Thereafter, the temperature drops first almost vertically to $\sim 2000$ K and then 
gradually to $\sim 300\ {\rm K}$ by ${\rm H}_2$ cooling.
In the cases with $J_{21} \geq 0.1$, without HD cooling, 
the gas cools only to the minimum temperature $\sim 200\ {\rm K}$ by H$_2$ cooling 
at $\sim 10^{3}\mbox{-}10^{4}\ {\rm cm^{-3}}$ as in the case of Pop III.1 star formation.
Namely, star formation proceeds in the H$_2$-Pop III mode. 
In this mode, the clouds fragment into rather massive clumps of 
$\sim 10^{3}\ {\rm M}_{\odot}$.
Note that, in the high density regime ($\ga 10^{5}\ {\rm cm^{-3}}$), 
all the evolutionary tracks converge to either of the two tracks, 
one by HD cooling for $J_{21} \leq 0.01$ or the other 
by ${\rm H}_2$ cooling for $J_{21} \geq 0.1$. 

\begin{figure}
\begin{center}
\includegraphics[scale=1.0]{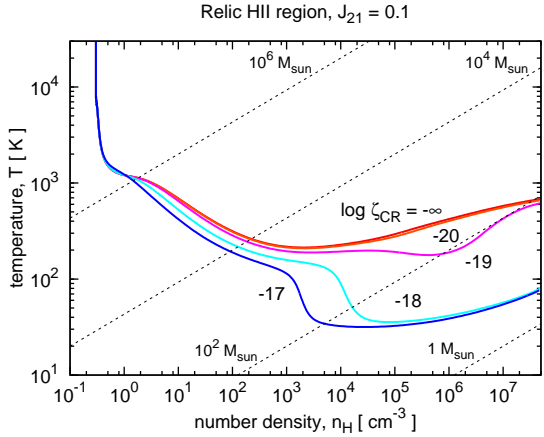}
\caption{CR effect on thermal evolution of clouds in relic HII regions.
The FUV intensity is fixed to $J_{21} = 0.1$, and the initial density to $n_{\rm H, 0} = 0.3\ {\rm cm}^{-3}$.
Different lines correspond to the different CR intensities: $\log \zeta_{\rm CR} ({\rm s}^{-1})= - \infty, -20, -19, -18$, and $-17$. The thin dashed lines indicate the constant Jeans masses.}
\label{fig:h2_nT_J01}
\end{center}
\end{figure}

\begin{figure}
\begin{center}
\includegraphics[scale=1.0]{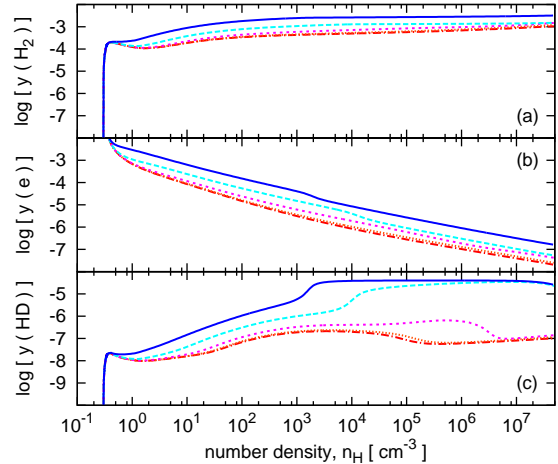}
\caption{(a) H$_2$, (b) $e$, and (c) HD fractions for the same clouds as in Fig. \ref{fig:h2_nT_J01}.
In each panel, the different lines correspond to the different values of 
CR intensity: 
$\log \zeta_{\rm CR}({\rm s}^{-1}) = - \infty$ (dot-dashed), $-20$ (dotted), $-19$ (short-dashed), $-18$ (long-dashed), and $-17$ (solid).}
\label{fig:h2_nT_J01_chem}
\end{center}
\end{figure}

Next, we see the results with CR irradiation on cloud evolution,  
as shown in Figs. \ref{fig:h2_nT_J01} and \ref{fig:h2_nT_J01_chem}, 
for the cases with FUV intensity $J_{21} = 0.1$.
The different lines in the Figures show different values of the CR intensity: 
$\log \zeta_{\rm CR} ({\rm s}^{-1})= - \infty$ (i.e., $\zeta_{\rm CR}=0$), 
$-20, -19, -18$ and $-17$.
Recall that, with $J_{21}=0.1$, HD cooling is suppressed without CR irradiation.
As seen in Fig. \ref{fig:h2_nT_J01}, with higher $\zeta_{\rm CR}$, the temperature becomes lower.
This is because the elevated ionization degree by CR ionization 
causes more efficient H$_2$ formation and cooling (see Fig. \ref{fig:h2_nT_J01_chem}), 
which outweighs the effect of CR ionization heating. 
For $\log \zeta_{\rm CR} \geq -18$, the temperature falls below 
150 K, where the HD fraction increases significantly, owing to 
the enhanced ${\rm H}_2$ cooling~(solid and long-dashed lines in Fig. \ref{fig:h2_nT_J01_chem}c).
By HD cooling, the temperature further drops to $\sim 30\ {\rm K}$, and HD-Pop III star formation 
is realized in these cases.
In this way, CR ionization can compensate the negative feedback of FUV photodissociation.

Since in our model for the free-falling cloud, the collapse equation~(eq. \ref{eq:drho_dt}) is decoupled from the energy equation~(eq. \ref{eq:energy}), the collapse time becomes the same for the same initial density and is obtained as $t_{\rm collapse} \sim 200$ Myr for $n_{{\rm H},0} = 0.3\ {\rm cm}^{-3}$.
The Hubble time is estimated as $t_{\rm H}(z) = 500 \left((1+z)/10\right)^{-3/2}\ {\rm Myr}$ and this is roughly equal to the galaxy merger time. We find that the collapse timescale estimated above is smaller than the Hubble time for $z < 15$.
Moreover, in the early universe we consider, the halo does not have an angular momentum large enough to form a rotation-supported disk which is comparable to the protogalaxy scale~(Abel et al. 2002).
Thus, the gas cloud in a relic HII region may remain isolated without being affected by the protogalaxy rotation and merger, and star formation there proceeds within the local Hubble time.

\subsubsection{Critical FUV intensity for HD Cooling}\label{subsec:h2_zeta_Jcrit}
\begin{figure}
\begin{center}
\includegraphics[scale=1.0]{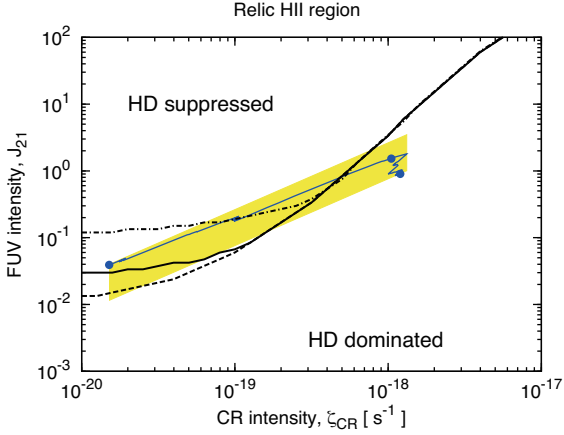}
\caption{The critical FUV intensity $J_{21, {\rm crit}}$ for HD cooling as a function of the CR intensity $\zeta_{\rm CR}$.
Each curve corresponds to the different initial densities: $n_{\rm H, 0} = 0.03$ (dashed), 0.3  (solid) and 3 (dot-dashed)$\ {\rm cm}^{-3}$.
Below $J_{21, {\rm crit}}$, HD dominates the cooling in the course of cloud collapse.
The shaded region shows the background FUV and CR intensities estimated from Eqs. \eqref{eq:J_theory} and \eqref{eq:zeta_theory} with $6 \la z \la 15$ and $10^{-4} \lesssim \Psi_{\ast, \rm II}(z) / {\rm M}_{\odot} {\rm yr}^{-1} {\rm Mpc}^{-3} \lesssim 3 \times 10^{-2}$.
The blue thin-solid line shows the evolution of the background intensities evaluated using the Pop II SFR $\Psi_{\ast, \rm II}(z)$ of Johnson et al. (2013).
The blue points refer to values at $z = 15, 10$ and $6$ from left to right.}
\label{fig:h2_zeta_Jcrit}
\end{center}
\end{figure}

We have learned that with the FUV intensity exceeding a critical value 
$J_{21, {\rm crit}}$, 
HD cooling is totally suppressed even in initially ionized clouds.
In addition, the value of the critical FUV intensity is elevated under CR irradiation.
In Fig. \ref{fig:h2_zeta_Jcrit}, the critical FUV intensity $J_{21, {\rm crit}}$ is plotted  
as a function of $\zeta_{\rm CR}$ for initial densities $n_{\rm H, 0} = 0.03$ (dashed), 0.3  (solid) and 3 (dot-dashed)$\ {\rm cm}^{-3}$.
In finding the value of $J_{21, {\rm crit}}$, we judge 
HD cooling is effective if the following two conditions are satisfied:
(i) the minimum temperature at $> 10^{5}\ {\rm cm^{-3}}$ is less than 100\ K, 
and (ii) the HD-cooling rate exceeds a half of the total (i.e. the sum of compressional and CR) heating rate at some moment.
These two conditions are imposed to discriminate whether the high-density temperature tracks are 
either those by H$_2$ cooling or by HD cooling.   
In the fiducial case of $n_{\rm H, 0}=0.3\ {\rm cm^{-3}}$ (solid line), 
the critical intensity is $J_{21, {\rm crit}} = 0.03$ for the negligible CR intensity 
($\zeta_{\rm CR} \sim 10^{-20}\ {\rm s}^{-1}$), in accord with previous 
studies~\citep[Yoshida et al. 2007b;][]{Wolcott2011}.
The critical intensity $J_{21, {\rm crit}}$ increases with the CR strength $\zeta_{\rm CR}$,
since strong enough CR irradiation enables efficient HD cooling even under an intense FUV field.

In the weak CR regime $\zeta_{\rm CR} \la 3 \times 10^{-19}\ {\rm s}^{-1}$, the electrons present initially, rather than those produced by CR ionization, work as catalysts for the H$_2$ formation reaction.
Abundant H$_2$ is formed if the self-shielding against the FUV field becomes effective before the ionization degree 
is reduced significantly by recombination.
Since the self-shielding is more effective for larger column densities and thus 
for larger $n_{\rm H, 0}$, more H$_2$ is formed under the same FUV intensity. 
This results in higher $J_{21, {\rm crit}}$ for higher $n_{\rm H, 0}$.
On the other hand, in the strong CR regime $\zeta_{\rm CR}$ $(\gtrsim 3 \times 10^{-19}\ {\rm s}^{-1})$, CR ionization provides the necessary electrons for abundant H$_2$ formation.
Moreover, with a FUV field as high as $J_{21, {\rm crit}}$, the shielding against the FUV field becomes effective
and ${\rm H}_2$ formation occurs only after the density is significantly increased from the initial value.
Therefore, the evolutionary track and critical FUV intensity $J_{21, {\rm crit}}$ do not depend on the initial density. 

\subsection{Shock-compressed Gas}\label{subsec:shock_result}
Here, we present the results for a shock-compressed gas in the two cases of the shock velocity: 
(i) $v_0 = 20\ {\rm km}\ {\rm s}^{-1}$ (virialization shock) and 
(ii) $100\ {\rm km}\ {\rm s}^{-1}$ (SN shock), respectively.

\subsubsection{Virialization Shock}\label{subsec:virial_result}
\begin{figure}
\begin{center}
\includegraphics[scale=1.0]{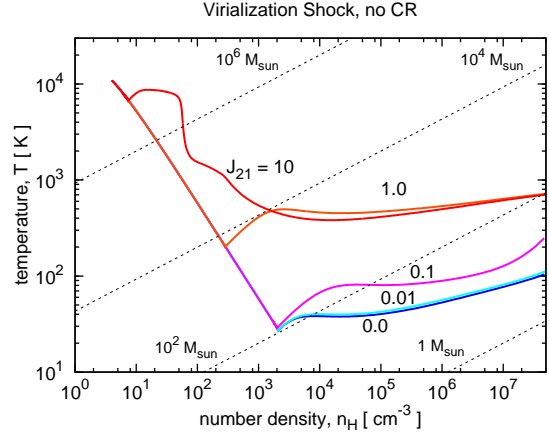}
\caption{FUV effect on thermal evolution of shock-compressed clouds by virialization.
No CR irradiation is considered. The initial density and shock velocity are taken as fiducial values: $n_{\rm H, 0} = 1.0\ {\rm cm}^{-3}$ and $v_0 = 20\ {\rm km}\ {\rm s}^{-1}$. Different lines correspond to the different FUV intensities: $J_{21}=0.0, 0.01, 0.1, 1.0$, and $10$.}
\label{fig:shock20_nT_CR0}
\end{center}
\end{figure}

\begin{figure}
\begin{center}
\includegraphics[scale=1.0]{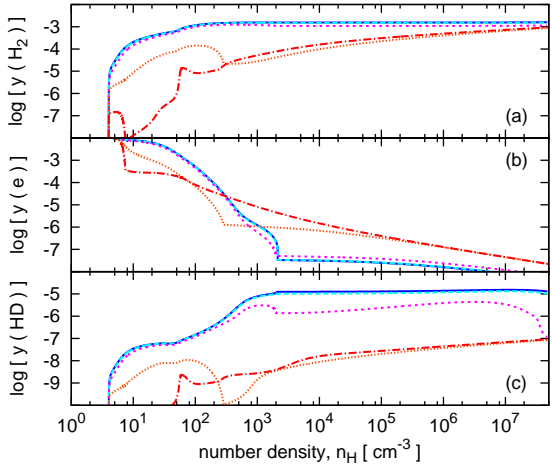}
\caption{Fractions of (a) H$_2$, (b) $e$, and (c) HD in the same clouds as in Fig. \ref{fig:shock20_nT_CR0}.
Different lines correspond to the different FUV intensities: $J_{21}=0.0$ (solid), $0.01$ (long-dashed), $0.1$ (short-dashed), $1.0$ (dotted), and $10$ (dot-dashed).}
\label{fig:shock20_nT_CR0_chem}
\end{center}
\end{figure}

First, to see the FUV effect on the gas compressed by the virialization shock 
($v_{0}=20\ {\rm km}\ {\rm s}^{-1}$), the results are shown 
for the cases without CR irradiation in Figs. \ref{fig:shock20_nT_CR0} and \ref{fig:shock20_nT_CR0_chem}.
Just behind the shock front, the gas is heated to $\gtrsim 10^4\ {\rm K}$ and then 
cools immediately to $\sim 8000\ {\rm K}$ by Ly-$\alpha$ cooling.
With the negligible or weak FUV field ($J_{21} \la 0.01$), H$_2$ forms abundantly owing to the high ionization degree 
($\sim 0.01$) in the post-shock gas (solid and long-dashed lines in Fig. \ref{fig:shock20_nT_CR0_chem}a, b). 
As in the relic HII region, the gas enters the HD formation regime ($\la 150$ K) 
by H$_2$ cooling (solid and long-dashed lines in Fig. \ref{fig:shock20_nT_CR0_chem}c), 
and the temperature further decreases to $\sim 30\ \rm K$ by HD cooling.
As long as the cooling time is shorter than the free-fall time, the post-shock gas is compressed  
almost isobarically~\citep[e.g.,][]{Shapiro1987, Yamada1998}.
However, below $\sim 30\ {\rm K}$, HD cooling is not efficient anymore, 
and the cooling time exceeds the free-fall time.
When the sufficient mass for gravitational instability accumulates 
in the dense layer, it fragments into clumps of approximately the Jeans mass 
$\sim 100\ {\rm M}_{\odot}$, which then continue collapsing by HD cooling
in the free-fall manner.
This late-time temperature evolution is along the HD-cooling track, like the free-falling clouds in relic HII regions~($J_{21}=0.0$ in Fig. \ref{fig:h2_nT_CR0}).
For $J_{21} = 0.1$, the isobaric contraction proceeds in the same way until $\sim 30\ \rm K$ 
as with the weaker FUV case.
However, during the dense gas layer stays at $\sim 30\ \rm K$ until the Jeans instability sets in, the HD fraction is significantly reduced by electron recombination and H$_2$/HD photodissociation~(short-dashed lines in Fig. \ref{fig:shock20_nT_CR0_chem}c).
As a result, HD cooling does not become important in the subsequent free-fall phase and 
temperature evolution eventually converges to the H$_2$-cooling track. 
For $J_{21} \geq 1.0$, strong FUV radiation completely quenches HD cooling
and massive clumps of $\sim 10^4\ {\rm M}_{\odot}$ will be formed.
Especially, in the case of $J_{21} = 10$, fragmentation takes place early at 
$\sim 8000\ \rm K$. The subsequent free-fall evolution is similar to that of the free-falling clouds 
in relic HII regions with $J_{21} \geq 1.0$ (Fig. \ref{fig:h2_nT_CR0}).

\begin{figure}
\begin{center}
\includegraphics[scale=1.0]{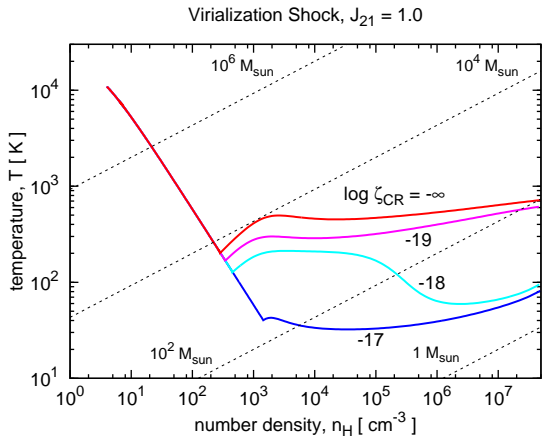}
\caption{CR effect on the thermal evolution of shock-compressed clouds by virialization.
The FUV intensity is fixed to $J_{21} = 1.0$.
The initial density and velocity are the fiducial values: $n_{\rm H, 0} = 1.0\ {\rm cm}^{-3}$ and $v_0 = 20\ {\rm km}\ {\rm s}^{-1}$.
The different lines correspond to the different CR intensities: $\log \zeta_{\rm CR}({\rm s}^{-1}) = - \infty, -19, -18$, and $-17$.}
\label{fig:shock20_nT_J1}
\end{center}
\end{figure}

\begin{figure}
\begin{center}
\includegraphics[scale=1.0]{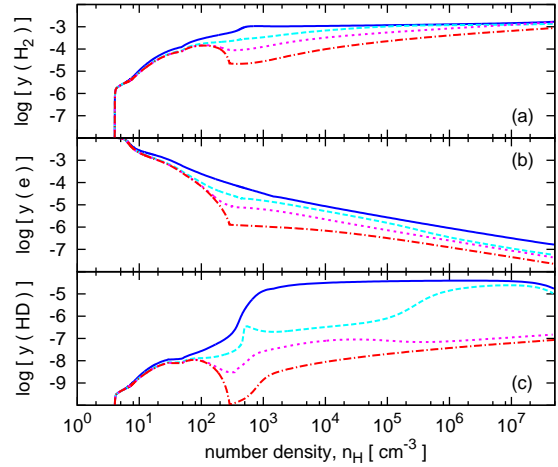}
\caption{Fractions of (a) H$_2$, (b) $e$, and (c) HD in the same clouds as in Fig. \ref{fig:shock20_nT_J1}.
The different lines correspond to the different CR intensities: $\log \zeta_{\rm CR}({\rm s}^{-1}) = - \infty$ (dot-dashed), $-19$ (short-dashed), $-18$ (long-dashed), and $-17$ (solid).}
\label{fig:shock20_nT_J1_chem}
\end{center}
\end{figure}

Next, we see the CR effect for the case with FUV intensity $J_{21}=1.0$. 
The results are shown in Figs. \ref{fig:shock20_nT_J1} and \ref{fig:shock20_nT_J1_chem}.
Recall that, with $J_{21} = 1.0$, HD cooling is suppressed in the absence of CR irradiation.
As in the case of the relic HII region, more ${\rm H}_2$ is formed by CR ionization~(Fig. \ref{fig:shock20_nT_J1_chem}a, b) and the temperature becomes lower by the enhanced H$_2$ cooling.
With the CR strength exceeding $\zeta_{\rm CR} = 10^{-18}\ {\rm s}^{-1}$, most deuterium is converted 
into HD~(long-dashed and solid lines in Fig. \ref{fig:shock20_nT_J1_chem}c), and temperature falls 
below $\la 100$ K. Thermal evolution finally converges to the HD-cooling track in the free-falling phase.

In our model for the shock-compressed gas, the contraction equations~(eqs. \ref{eq:rho_v}, \ref{eq:rho_v^2}) are coupled with the energy equation~(eq. \ref{eq:energy}) before the self gravity becomes effective, so that the timescale of the isobaric contraction phase differs among the lines. 
However, the evolution time after self-gravity becomes effective is almost the same and the time to reach $n=10^5\ {\rm cm}^{-3}$ is $\sim 50$ Myr for the structure-formation shock, which is by an order of magnitude smaller than the local Hubble time $t_{\rm H}(z) = 500 \left((1+z)/10\right)^{-3/2}\ {\rm Myr}$ for $z < 15$.
Thus, the shock-compressed gas in structure formation may be unaffected by the protogalaxy rotation and merger, and star formation there proceeds within the local Hubble time.

\subsubsection{Supernova Shock}\label{subsec:SN_result}
\begin{figure}
\begin{center}
\includegraphics[scale=1.0]{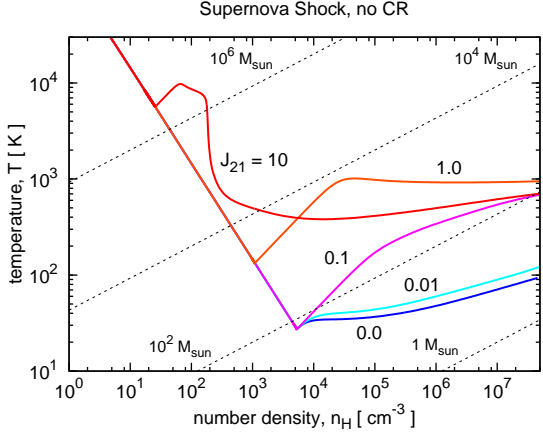}
\caption{FUV effect on the thermal evolution of shock-compressed clouds by SNe.
No CR irradiation is considered. 
The fiducial initial conditions are adopted: $n_{\rm H, 0} = 0.1\ {\rm cm}^{-3}$ and $v_0 = 100\ {\rm km}\ {\rm s}^{-1}$.
The different lines correspond to the different FUV intensities: $J_{21}=0.0, 0.01, 0.1, 1.0$, and $10$.}
\label{fig:shock100_nT_CR0}
\end{center}
\end{figure}

In Fig. \ref{fig:shock100_nT_CR0}, we show the FUV effect on the shock-compressed gas in an SN explosion 
($v_0=100\ {\rm km}\ {\rm s}^{-1}$) for cases without CR irradiation.
We find that, despite the different initial conditions, the evolutionary tracks are similar 
to those in the virialization shock with the same FUV intensity~(see Fig. \ref{fig:shock20_nT_CR0}).
This is because, although the initial density $n_{{\rm H}, 0} = 0.1\ {\rm cm}^{-3}$ is ten times lower 
in the SN case, the post-shock temperature is 25 times higher and thus the pressure in the isobaric 
evolution phase differs only by a factor of 2.5~(see Fig. \ref{fig:shock20_nT_CR0}).
We also find that the chemical compositions vary in a similar way as in the case of virialization 
shock~(see Fig. \ref{fig:shock20_nT_CR0_chem}), 
except that the ionization degree reaches $\sim 1$ due to the very high temperature 
$\sim 10^5\ {\rm K}$ behind the SN shock front.

\begin{figure}
\begin{center}
\includegraphics[scale=1.0]{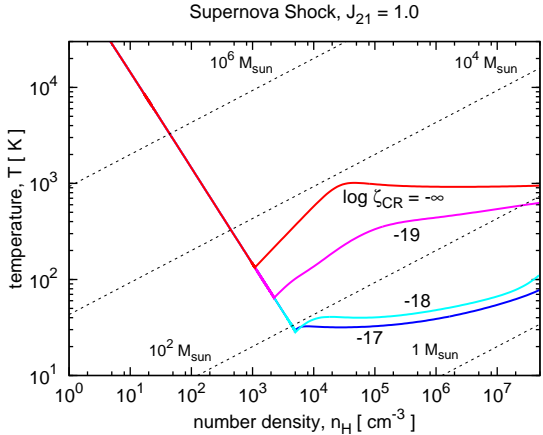}
\caption{CR effect on the same clouds as in Fig. \ref{fig:shock100_nT_CR0}.
The FUV intensity is fixed to $J_{21} = 1.0$.
The different lines correspond to the different CR intensities: $\log \zeta_{\rm CR}({\rm s}^{-1}) = - \infty, -19, -18$, and $-17$.}
\label{fig:shock100_nT_J1}
\end{center}
\end{figure}

The same trend holds also in the cases with CR irradiation, and the evolutionary tracks are similar 
to those in the virialization shock with the same CR intensity~(see Figs. \ref{fig:shock20_nT_J1} and \ref{fig:shock100_nT_J1}).

For the SN-shock case, the evolution time after self-gravity becomes effective is almost the same and the time to reach $n=10^5\ {\rm cm}^{-3}$ is calculated as $\sim 250$ Myr, which is much longer than the lifetime of an SN remnant $\sim 1$ Myr.
Thus, the SN shock will dissolve into the ISM before it collects enough materials to trigger fragmentation and star formation in the post-shock region.
This result is consistent with Chiaki et al. (2013), where they find that the SN-shock-compressed gas can collapse within the lifetime of the SN remnant for the ISM density $\gtrsim 1\ {\rm cm}^{-3}$ and $E_{\rm SN} = 10^{51}$ erg.
However, we discuss the evolution of the SN-shock-compressed gas for comparison with that of structure formation.

\subsubsection{Critical FUV intensity for HD Cooling}\label{subsec:sh_zeta_Jcrit}
\begin{figure}
\begin{center}
\includegraphics[scale=1.0]{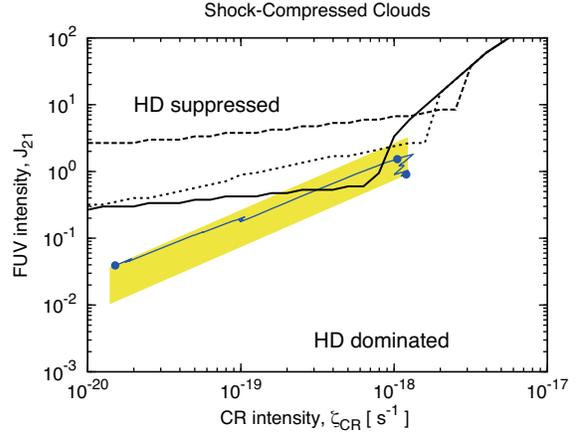}
\caption{The critical FUV intensity $J_{21, {\rm crit}}$ for HD cooling as a function of the CR intensity $\zeta_{\rm CR}$ in shock-compressed clouds~(see also Fig. \ref{fig:h2_zeta_Jcrit}).
The solid and dashed lines correspond to the virialization shock with the fiducial case, $(v_0, n_{\rm H, 0}) = (20, 1.0)$, and the high density case, $(v_0, n_{\rm H, 0}) = (20, 10)$, respectively.
The dot-dashed line shows the SN-shock case, $(v_0, n_{\rm H, 0}) = (100, 0.1)$.}
\label{fig:sh_zeta_Jcrit}
\end{center}
\end{figure}

As in \S \ref{subsec:h2_zeta_Jcrit}, 
the critical FUV intensity for HD cooling $J_{21, {\rm crit}}$ is calculated 
for a shock-compressed gas
and is presented in Fig. \ref{fig:sh_zeta_Jcrit} as a function of the CR intensity $\zeta_{\rm CR}$.
The solid and dashed lines are for the virialization shock in the fiducial case 
$(v_0, n_{\rm H, 0}) = (20, 1.0)$ and in the high density case 
$(v_0, n_{\rm H, 0}) = (20, 10)$, respectively, and 
the dot-dashed line indicates the SN shock case $(v_0, n_{\rm H, 0}) = (100, 0.1)$.

From Fig. \ref{fig:sh_zeta_Jcrit}, we see that for $\log \zeta_{\rm CR} \lesssim -18$, the critical value $J_{21, {\rm crit}}$ becomes higher for a higher initial density $n_{\rm H, 0}$ or higher shock velocity $v_0$.
The reason is the same as in the case of a relic HII region:
in the weak CR regime, abundant H$_2$/HD formation is caused by the
high post-shock ionization degree, and thus the prompt FUV shielding 
before significant recombination is needed.
This can be realized more easily with a higher initial density $n_{\rm H, 0}$ or higher shock velocity $v_0$,
since the column density at $\sim 10^4$ K, below which 
recombination proceeds significantly, is higher in these cases.
On the other hand, for $\log \zeta_{\rm CR} \gtrsim -18$, 
the critical value $J_{21, {\rm crit}}$ is solely determined by the CR strength $\zeta_{\rm CR}$ and 
becomes independent of the initial condition $(n_{\rm H, 0}, v_0)$.
Again the reason is the same as in the relic-HII-region case: 
the electrons from CR ionization, rather than those present initially, now work 
as catalysts for H$_2$ formation and thus the H$_2$ formation rate does not depend on the 
initial parameters.

The comparison between Figs. \ref{fig:h2_zeta_Jcrit} and \ref{fig:sh_zeta_Jcrit} tells us
that in the weak CR regime, $\log \zeta_{\rm CR} \la -19$, the critical intensity is $\sim 10$ times higher 
for the shock-compressed gas ($\log J_{21,{\rm crit}}=-0.5...0.5$) than in the relic HII region 
($\log J_{21,{\rm crit}}=-2...-1$).
In this case, for abundant H$_2$/HD formation, H$_2$ should be promptly shielded against the FUV field 
before the electrons present initially recombine significantly.
This condition can be more easily satisfied in the shock-compressed gas than in the relic HII region.
The reason is that the shock-compressed gas evolves isobarically~(see Fig. \ref{fig:shock20_nT_CR0}) rather 
than isochorically as in the relic HII region~(see Fig. \ref{fig:h2_nT_CR0}), which leads to the larger column density in the shock-compressed gas.
In addition, shocks tend to occur in denser environments~(e.g., cold-accretion flows) than 
relic HII regions.
Consequently, in the shock-compressed gas, H$_2$ is shielded from the FUV field more effectively, and the critical FUV intensity becomes higher than that in the relic HII region.

\section{Estimate for background intensities}\label{sec:estimate}
We have seen that Pop III.2 star formation proceeds through the HD-cooling track, 
i.e., in the HD-Pop III star formation mode, under a weak FUV field or strong enough CR irradiation.
In this section, we estimate the intensities of the FUV and CR backgrounds as a function of redshift.
Then, by comparing them with the critical value $J_{21, {\rm crit}}$ obtained in \S \ref{sec:result}, 
we discuss whether HD-Pop III star formation is realized at the epoch of galaxy formation~($z \lesssim 15$).

At this epoch, metal enrichment in an average galaxy may have proceeded to 
the level for Pop II star formation to occur~(e.g., Wise et al. 2012; Johnson et al. 2013).
We thus consider massive Pop II stars and their SN remnants as the plausible sources of the FUV and CR backgrounds, 
respectively. 
Then, the intensity of the FUV background is estimated as~\citep{Greif2006}
\begin{equation}
J_{\rm LW, bg}(z) = \frac{hc}{4\pi m_{\rm H}} \frac{\bar{\nu}}{\Delta \nu} \eta_{\rm LW} \rho_*(z) (1+z)^3,
\label{eq:BFUV_1}
\end{equation}
where $h$ is the Planck constant, $c$ the speed of light, $\bar{\nu}$ the average frequency of the LW band ($h \bar{\nu} = 12.4$ eV), $\Delta \nu$ the LW band width ($h \Delta \nu = 2.4$ eV), $\eta_{\rm LW} = 4000$ the number of FUV photons emitted per stellar baryon, and $\rho_*(z)$ the stellar mass density at $z$.
Johnson et al. (2013) evaluated $\rho_*(z)$ by the relation 
$\rho_*(z) = \Psi_{\ast, \rm II}(z)\ t_{\ast}$, 
where $\Psi_{\ast, \rm II}(z)$ is the Pop II SFR density, and $t_{\ast}=5\ {\rm Myr}$ the lifetime of a massive star.
Then, Eq. \eqref{eq:BFUV_1} can be rewritten as
\begin{equation}
J_{21, \rm bg}(z) \sim 0.1 \left(\frac{\Psi_{\ast, \rm II}(z)}{10^{-3}\ {\rm M}_{\odot}\ {\rm yr}^{-1}\ {\rm Mpc}^{-3}}\right) \left(\frac{1+z}{11}\right)^{3}.
\label{eq:J_theory}
\end{equation}
Following \cite{Stacy2007}, the energy density of the background CRs $U_{\rm CR}(z)$ is estimated from
\begin{equation}
U_{\rm CR}(z) \sim p_{\rm CR} E_{\rm SN} f_{\rm SN} \Psi_{\ast, \rm II}(z) t_{\rm H}(z) (1+z)^3,
\label{eq:U_theory}
\end{equation}
\begin{equation}
\sim 4 \times 10^{-17} \left(\frac{\Psi_{\ast, \rm II}(z)}{10^{-3}\ {\rm M}_{\odot} {\rm yr}^{-1}\ {\rm Mpc}^{-3}}\right)\left(\frac{1+z}{11}\right)^{\frac{3}{2}}{\rm erg}\ {\rm cm}^{-3}, \notag
\end{equation}
where $E_{\rm SN}=10^{51}\ {\rm erg}$ is the SN energy, $p_{\rm CR}=0.1$ the 
efficiency of $E_{\rm SN}$ converted to CR acceleration, $f_{\rm SN}$ the number of SNe per 
unit mass of stars formed, and $t_{\rm H}(z)$ the Hubble time at $z$.
$f_{\rm SN}$ is obtained as $\sim 2 \times 10^{-2}\ {\rm M}_{\odot}^{-1}$ 
assuming that massive stars with $\gtrsim 8\ {\rm M}_{\odot}$ explode as SNe 
for the Salpeter IMF.
The CR ionization rate of a hydrogen atom $\zeta_{\rm CR}$ is related with $U_{\rm CR}$ as 
\citep{Inayoshi2011}:
\begin{equation}
\zeta_{\rm CR} = 1.7 \times 10^{-18} \left(\frac{U_{\rm CR}}{10^{-15}\ {\rm erg}\ {\rm cm}^{-3}}\right)\ {\rm s}^{-1}.
\label{eq:zeta_U}
\end{equation}
Combining Eqs. \eqref{eq:U_theory} and \eqref{eq:zeta_U},
\begin{equation}
\zeta_{\rm CR}(z) \sim 6.8 \times 10^{-20}\left(\frac{\Psi_{\ast, \rm II}(z)}{10^{-3}{\rm M}_{\odot}{\rm yr}^{-1}{\rm Mpc}^{-3}}\right)\left(\frac{1+z}{11}\right)^{\frac{3}{2}}{\rm s}^{-1}.
\label{eq:zeta_theory}
\end{equation}

The high-$z$ SFR density has been evaluated theoretically by some authors~(e.g., Tornatore et al. 2007; Trenti \& Stiavelli 2009; Johnson et al. 2013).
According to Johnson et al. (2013), $\Psi_{\ast, \rm II}(z)$ increases almost monotonically from $10^{-4}$ to $3 \times 10^{-2}\ {\rm M}_{\odot} {\rm yr}^{-1} {\rm Mpc}^{-3}$ as $z$ decreases from $15$ to $6$.
The blue thin-solid lines in Figs. \ref{fig:h2_zeta_Jcrit} and \ref{fig:sh_zeta_Jcrit} show the evolution of $J_{21, \rm bg}(z)$ and $\zeta_{\rm CR}(z)$ evaluated from Eqs. \eqref{eq:J_theory}, \eqref{eq:zeta_theory} and $\Psi_{\ast, \rm II}(z)$ of Johnson et al. (2013).
The background intensities move from left to right along the line, and the blue points indicate the values 
at $z = 15, 10$ and $6$, respectively.
Although the high-$z$ SFR density is still uncertain, most studies concluded $10^{-4} \lesssim \Psi_{\ast, \rm II}(z)/{\rm M}_{\odot} {\rm yr}^{-1} {\rm Mpc}^{-3} \lesssim 3 \times10^{-2}$ for $6 \lesssim z \lesssim 15$~(e.g., Tornatore et al. 2007; Trenti \& Stiavelli 2009; Johnson et al. 2013).
The shaded regions in Figs. \ref{fig:h2_zeta_Jcrit} and \ref{fig:sh_zeta_Jcrit} indicate the range of $J_{21, \rm bg}(z)$ and $\zeta_{\rm CR}(z)$ corresponding to this parameter range.

First, we examine the case of free-falling clouds in relic HII regions~(Fig. \ref{fig:h2_zeta_Jcrit}).
For initial densities lower than the fiducial case $n_{\rm H, 0}=0.3\ {\rm cm^{-3}}$~(black solid line),
the critical FUV intensity for HD cooling $J_{21, {\rm crit}}$ is below the background level $J_{21, \rm bg}$ at $z \gtrsim 10$, and HD cooling is suppressed in Pop III.2 star formation.
As the background CR intensity $\zeta_{\rm CR}$ increases with cosmic star formation, the critical intensity 
$J_{21, {\rm crit}}$ rises and eventually exceeds the background level $J_{21, \rm bg}$ at $z \lesssim 10$.
Below this redshift, HD cooling becomes efficient and HD-Pop III star formation occurs if some primordial environments still survive.
Although in a very dense HII region~(dot-dashed line), the critical intensity $J_{21, {\rm crit}}$ is above the background level $J_{21, \rm bg}$ even at $z \gtrsim 10$, such dense conditions appear to be hardly realized~(Kitayama et al. 2004; Whalen et al. 2004, Yoshida et al. 2007a,Abel2007).
We conclude that, in relic HII regions, HD-Pop III star formation is possible only at $z \lesssim 10$, where the intensity of the CR background is modest.

On the other hand, for the shock-compressed gas (Fig. \ref{fig:sh_zeta_Jcrit}), the critical value $J_{21, {\rm crit}}$ exceeds the background level $J_{21, \rm bg}$ for most of 
the redshift we consider ($6 \la z \la 15$), and thus HD-Pop III star formation occurs. 
Note that the pre-shock densities we adopt in Fig. \ref{fig:sh_zeta_Jcrit} are rather low, 
and our results may give the lower bound of the critical FUV intensity $J_{21, {\rm crit}}$.
In fact, in the mass assembly process of first galaxies, cold-streaming shocks occur at the regions as dense as $n_{{\rm H}, 0} = 1\mbox{-}10^3\ {\rm cm}^{-3}$~(Wise \& Abel 2007; Wise et al. 2008).
In such a dense shock environment, HD-Pop III star formation would be more common than the cases we have studied. 
In conclusion, the HD-Pop III mode is the main mode of Pop III.2 star formation in the cloud compressed either by a virialization or SN shock. 

\section{Discussion}\label{sec:discussion}
So far, we have considered that the background FUV photons are uniformly distributed in the universe.
In reality, they have some spatial fluctuations caused by local stellar sources, like star-forming galaxies~(Dijkstra et al. 2008; Ahn et al. 2009; Agarwal et al. 2012; Johnson et al. 2013).
In the primordial clouds irradiated with the strong FUV field as $J_{21} > 3$, HD formation is suppressed even in the shock-compressed gas unless the initial density is high ($10\ {\rm cm}^{-3}$) or the CR intensity is strong 
($> 10^{-18}\ {\rm s}^{-1}$: see Fig. 12).
However, according to Johnson et al. (2013), only $\la 10$ percent of the dense primordial cloud with $n_{\rm H} > 1\ {\rm cm}^{-3}$ is exposed to such a strong FUV field as $J_{21} > 3$.
Thus, our analysis with the uniform background FUV field is enough to discuss the major character of primordial star formation and the spatial fluctuations of it do not affect our main conclusions.

We also assume that CRs accelerated in SN remnants escape freely from their host galaxies and fill the universe homogeneously.
However, in reality, they may be trapped in the host galaxy by the interaction with the magnetic field, and propagate diffusively through the ISM of the galaxy~\citep{Longair2011}.
On the way of propagation, CRs also lose their kinetic energy by ionizing the neutral ISM.
For the efficient CR escape into the IGM, the CR escape time by diffusion $\tau_{\rm diff}$ should be shorter than both the ionization loss time $\tau_{\rm ion}$ and the age of the system $\tau_{\rm age}$, which is roughly equal to the age of the universe. Here, $\tau_{\rm diff}$ and $\tau_{\rm ion}$ depends on the kinetic energy of a CR particle $E$  and the redshift $z$.
\cite{Rollinde2008} estimated the above timescales for CRs with $E=30$ MeV and discussed the CR escape efficiency during the cosmic structure formation.
Here, we extend their arguments by examining the $E$ dependence of the timescales.
Following \cite{Stacy2007}, we assume that the energy spectrum of the CR number density has a power-law shape $dn_{\rm CR}/d E \propto E^{-x}$ with $E = 10^6\mbox{-}10^{15}$ eV.
We find that the inequality $\tau_{\rm diff}(E, z) < \tau_{\rm age}(z)$ holds for all the CR energy and redshift we consider~($6 \la z \la 15$).
On the other hand, the inequality $\tau_{\rm diff}(E, z) < \tau_{\rm ion}(E, z)$ holds only for high energy CRs with $E \ga 10^7$ eV at $6 \la z \la 15$.
Thus, the CR background is composed of CRs with $E \ga 10^7$ eV and its intensity varies with the spectrum index $x$.
In the standard Fermi acceleration theory, the spectrum index is given as $x = 2$~\citep[e.g.,][]{Bell1978}, and the background intensity is reduced only by 10\ \% due to the confinement of CRs with $E = 10^6\mbox{-}10^7$ eV.
On the other hand, if we adopt a steeper slope as $x = 3$ in~\cite{Rollinde2008}, it is reduced by an order of magnitude.
Even in this case, if a star-forming pristine gas resides in the same galaxy hosting SNe, it is exposed to the CR intensity in the original or even the enhanced level, because of the efficient confinement within the galaxy.
For more quantitive and definitive discussions, a detailed study of CR propagation in the ISM of high-$z$ galaxies is needed and is beyond the scope of this paper.

At the epoch of galaxy formation, high-mass X-ray binaries and miniquasars may be the plausible sources of high energy photons extending to the X-ray band~\citep{Glover2003}.
Since the photoionization cross section of neutral atoms generally decreases with frequency in the power-law way, the IGM becomes optically thin for X-ray photons with $\ga 1$ keV and the X-ray background develops at this epoch~(Haiman et al. 2000).
The injection of X-ray photons enhances the cooling efficiency of primordial clouds by indirectly promoting H$_2$ and HD formation through the photoionization of neutral atoms~\citep[e.g.,][]{Haiman1996a,Haiman2000,Glover2003,Machacek2003}.
Nonetheless, we did not include this X-ray feedback, and here we briefly discuss how it changes our results. 
Following \cite{Glover2003}, we assume that the X-ray background has a power-law spectrum
\begin{equation}
J_{\rm X}(\nu) = J_{{\rm X}, 21} \times 10^{-21} \left(\frac{\nu}{1\ {\rm keV}}\right)^{-1.5}\ {\rm erg}\ {\rm cm}^{-2}\ {\rm s}^{-1}\ {\rm Hz}^{-1},
\label{eq:J_X}
\end{equation}
where $J_{\rm X, 21} \equiv J_{1 \rm keV}/10^{-21}\ {\rm erg}\ {\rm cm}^{-2}\ {\rm s}^{-1}\ {\rm Hz}^{-1}$ and $J_{1 \rm keV}$ is the intensity at 1 keV.
Then, using Eqs. (17-19) in \cite{Inayoshi2011}, the ionization rate of H is calculated as
\begin{equation}
\zeta_{\rm X}^{\rm H} \sim 2 \times 10^{-15}J_{\rm X, 21}\ {\rm s}^{-1},
\label{eq:zeta_X_J_X}
\end{equation}
for the hydrogen column density $N_{\rm H} \lesssim 10^{22}\ {\rm cm}^{-2}$.
Since the background field derives from the stellar activity, the X-ray intensity is related with the CR energy density as
\begin{equation}
J_{{\rm X}, 21} \sim 10^{-3} \left(\frac{U_{\rm CR}}{10^{-15}\ {\rm erg}\ {\rm cm}^{-3}}\right),
\label{eq:J_X_u_cr}
\end{equation}
where we suppose that CR sources are Pop II clusters with the Salpeter IMF and the mass range $1 \mbox{-}100\ {\rm M}_{\odot}$~\citep{Inayoshi2011}\footnote{We adopt $E_{\rm SN} = 10^{51}$ erg as the energy of a Pop II SN instead of $E_{\rm SN} = 10^{52}$ erg in \cite{Inayoshi2011}.}.
Substituting Eq. \eqref{eq:zeta_U} into Eq. \eqref{eq:J_X_u_cr} and Eq. \eqref{eq:J_X_u_cr} into Eq. \eqref{eq:zeta_X_J_X}, the X-ray ionization rate is related with the CR ionization rate as
\begin{equation}
\zeta_{\rm X}^{\rm H} \sim \zeta_{\rm CR}.
\label{eq:zetaX_zetaCR}
\end{equation}
Thus, we find that, at each redshift, X-ray feedback changes the ionization rate in pristine clouds not by an order of magnitude, but by only up to a factor of two.
Note that this is the upper limit of the X-ray effect since X-ray is more effectively shielded than CRs.
Therefore, our main conclusion still remains valid even if we consider X-ray feedback.

As we have seen above, Pop III star formation at the epoch of galaxy formation is affected significantly by the injection of CRs.
Nonetheless, some previous studies calculated the Pop III SFR by considering the regulation process 
of star formation from FUV feedback alone, and the positive feedback from CRs is neglected~(e.g., Tornatore et al. 2007; Trenti \& Stiavelli 2009; de Souza et al. 2011; Johnson et al. 2013).
Future studies should take all these feedback into consideration.

We must admit that our treatment of the plane-parallel and steady shock  
is too simplistic to capture possible complex phenomena induced by e.g., 
the turbulence, rotation, thermal instability, and magnetic field.
For example, \cite{Inoue2008,Inoue2009} studied the effect of magnetic fields on the thermal evolution of a shock-compressed gas, using the two-dimensional MHD simulation. They found that after the short isobaric contraction, the magnetic pressure is amplified to be comparable to the ram pressure of the pre-shock materials, and thereafter the shock-compressed gas cools almost isochorically.
This implies that the shielding of H$_2$ against a FUV field becomes less effective and the resultant H$_2$ fraction is lowered in the presence of magnetic fields.
In this case, we expect that more massive clumps may form after the gas becomes gravitationally unstable and fragments, compared to the isobaric contraction case without magnetic fields.
However, their calculation did not follow the evolution until the self-gravity of the shocked layer becomes important.
They did not consider the gas in the primordial composition as well.
Thus, the above effects on our results should be investigated through the 3D MHD simulations in the future.
Another example is that our assumption of the time-independent shock velocity becomes invalid at least in the SN-shock, according to the one-dimensional numerical calculations~\citep{Machida2005, Nagakura2009, Chiaki2013}.
They showed that the SN-shock velocity becomes lower than $100\ {\rm km}\ {\rm s}^{-1}$ at $\sim 10^5$ yr after the explosion.
Moreover, at $\gtrsim 6 \times 10^5$ yr, the ram pressure of the ambient medium decreases below the pressure in the shell so that the shell re-expands and decreases its temperature and density in an almost adiabatic manner.
Although in this case, H$_2$ is produced efficiently to $\sim 10^{-3}$ and the temperature decreases to $\sim 200$ K by H$_2$ cooling, HD is  produced only to $\sim 10^{-6}$, which is $\sim 10$ times less than our results, in the adiabatic expansion phase owing to the low gas density.
In this case, the temperature decrease below $100$ K is not caused by HD cooling but by the adiabatic expansion of the shell~\citep{Nagakura2009}.
However, if the SN shock can collect enough materials to trigger gas fragmentation and contraction by self-gravity before dissolving into the ISM, HD formation becomes effective and HD cooling becomes dominated in the free-falling phase, as we can see from Fig. 6 in \cite{Machida2005} and Fig. 6 in \cite{Chiaki2013}.
Thus, whether this realistic evolution of the SN shock front significantly changes our results is not clear, and more detailed numerical calculations are needed for more concrete discussion.

\section{Conclusions}\label{sec:conclusion}
HD molecules form abundantly in the primordial gas with enhanced initial ionization and play a dominant role in gas cooling.
At the epoch of galaxy formation~($z \lesssim 15$), shocks occur ubiquitously associated with the mass assembly 
as well as the SN explosions.
In such a shock-compressed gas, the ionization degree jumps up owing to collisional ionization.
Other examples include the cloud in the relic HII region of a defunct Pop III star and 
that with modest CR irradiation.
However, even a low level of the FUV background suppresses HD formation/cooling via ${\rm H}_2$ photodissociation 
in cases without CR irradiation.
In this paper, we have examined the conditions for efficient HD cooling in primordial star formation by calculating the thermal and chemical evolution of a gas cloud under both FUV and CR irradiation.
We have obtained the critical FUV intensity $J_{21, {\rm crit}}$ for HD cooling as a function of the CR intensity $\zeta_{\rm CR}$, and compared it with the estimated background level $J_{21, \rm bg}$.
We have considered both cases of the free-falling cloud in a relic HII region and a shock-compressed gas.

At $z \ga 10$, the background CR intensity is estimated as $\sim 0.1\ \%$ of the Galactic value. In this case, the critical FUV intensity $J_{21, {\rm crit}}$ in relic HII regions is below the background level $J_{21, \rm bg}$ and HD cooling is suppressed there.
However, as the background CR intensity increases with the cosmic star formation, the critical intensity $J_{21, {\rm crit}}$ is also elevated and eventually exceeds the background level $J_{21, \rm bg}$ at $z \la 10$.
Below this redshift, HD cooling becomes efficient and HD-Pop III star formation proceeds 
if some primordial environments still survive.

On the other hand, the critical FUV intensity $J_{21, {\rm crit}}$ for the shock-compressed gas is $\sim 10$ times higher than that for the relic HII region even in the weak CR regime as $\sim 0.1\ \%$ of the Galactic value.
This is because the shock-compressed gas evolves isobarically~(i.e., gas cools with its density increasing), while the gas in a relic HII region evolves isochorically~(i.e., the gas cools at a constant density) due to the initial rapid cooling.
Moreover, shocks tend to occur in denser environments~(e.g., cold-accretion flows) than relic HII regions.
As a result, the shock-compressed gas is shielded from the FUV field more effectively, and has higher critical intensity.
Thus, for the shock-compressed gas, the critical value $J_{21, {\rm crit}}$ exceeds the background level $J_{21, \rm bg}$ 
and HD-Pop III star formation proceeds for most of the redshift we consider ($6 \la z \la 15$).

Our result suggests that HD-Pop III stars can be more common than previously considered and could be 
even the dominant population of Pop III stars.
H$_2$-Pop III stars are born in clumps as massive as $\sim 10^3\ {\rm M}_{\odot}$, but the final stellar mass is set to $\sim 100\ {\rm M}_{\odot}$ by the stellar radiative feedback onto the accretion 
flow~\citep{McKee2008, Hosokawa2011,Hirano2014}.  
On the other hand, HD-Pop III stars form in clumps of $\sim 100\ {\rm M}_{\odot}$, and the final mass is typically set to a few $10\ {\rm M}_{\odot}$ by the stellar feedback \citep{Hosokawa2012}.
These previous results may suggest that the star formation efficiency, i.e., the mass ratio of formed
 stars to the parent clouds, of HD-Pop III stars is several times higher
 than that of H2-Pop III stars. They also imply the possibility that HD-Pop III stars could be the majority of Pop III.2 stars.
Then, HD-Pop III stars of a few 10 ${\rm M}_{\odot}$ can be important sources of radiative feedback 
at the epoch of galaxy formation.
Moreover, according to the observations of the abundance pattens in the metal-poor halo stars in the Galaxy, the imprints of a pair-instability SN have not been discovered yet~\citep{Tumlinson2004, Frebel2009}.
Since stars with $10\mbox{-}40\ {\rm M}_{\odot}$ end their lives as core-collapse SNe~\citep{Heger2003}, HD-Pop III stars might be the major contributor to the metal enrichment in the ISM/IGM at that epoch, if there were no PISNe in the universe.

Recent discoveries of the almost pristine gas with metallicity $\lesssim 10^{-4}\ {\rm Z}_{\odot}$ 
in the damped Ly $\alpha$ systems at $z \sim 3$ and $7$~\citep{Fumagalli2011,Simcoe2012}
imply that metal enrichment in the IGM has proceeded quite inhomogeneously.
Theoretical models also show the inhomogeneous nature of metal enrichment in the IGM.
They predict that Pop III star formation can continue 
up to $z \sim 6$ or even lower, and evaluate the ratio of the Pop III SFR to the total SFR as $\sim 10^{-3}\mbox{-}10^{-2}$ ($\sim 10^{-4}$) at $z=10$ ($z=6$)~(Tornatore et al. 2007; Johnson et al. 2013).
This indicates that Pop III stars can form to some extent even at $z \lesssim 10$, and our study show that 
they are typically formed as HD-Pop III stars with a few $10\ {\rm M}_{\odot}$.
GRBs and SNe from these Pop III progenitors at $z \la 10$ will be detected with the current or 
future facilities and may provide us with information on the mass, IMF and SFR of these stars~(Nakauchi et al. 2012; Whalen et al. 2013; Tanaka et al. 2013).
Moreover, if the HD absorption lines are detected in the spectrum of a GRB afterglow which 
lacks metal absorption, HD is confirmed as the dominant coolant in that primordial gas~\citep{Inoue2007}.

\section*{Acknowledgements}
We thank the anonymous referee for helpful comments and improving the quality of this paper.
We also thank K. Kashiyama, T. Nakamura Y. Suwa, and K. Tanaka for fruitful discussions and comments.
This work is supported in part by the Grant-in-Aid from the Ministry
of Education, Culture, Sports, Science and Technology (MEXT) of Japan~(23-838 KI; 25287040 KO).


\begin{thebibliography}{99}

\bibitem[Abel, Bryan \& Norman(2002)]{Abel2002} Abel, T., Bryan, G.~L., 
\& Norman, M.~L.\ 2002, Science, 295, 93

\bibitem[Abel, Wise \& Bryan(2007)]{Abel2007} Abel, T., Wise, J.~H., 
\& Bryan, G.~L.\ 2007, ApJ, 659, L87

\bibitem[Agarwal et al.(2012)]{Agarwal2012} Agarwal, B., Khochfar, 
S., Johnson, J.~L., et al.\ 2012, MNRAS, 425, 2854

\bibitem[Ahn et al.(2009)]{Ahn2009} Ahn, K., Shapiro, P.~R., 
Iliev, I.~T., Mellema, G., \& Pen, U.-L.\ 2009, ApJ, 695, 1430

\bibitem[Bell(1978)]{Bell1978} Bell, A.~R.\ 1978, MNRAS, 182, 
147

\bibitem[Bromm, Coppi \& Larson(1999)]{Bromm1999} Bromm, V., Coppi, P.~S., 
\& Larson, R.~B.\ 1999, ApJ, 527, L5

\bibitem[Bromm, Coppi \& Larson(2002)]{Bromm2002} Bromm, V., Coppi, P.~S., 
\& Larson, R.~B.\ 2002, ApJ, 564, 23

\bibitem[Bromm, Yoshida \& Hernquist(2003)]{Bromm2003} Bromm, V., Yoshida, N., 
\& Hernquist, L.\ 2003, ApJ, 596, L135

\bibitem[Chiaki, Yoshida \& Kitayama(2013)]{Chiaki2013} Chiaki, G., Yoshida, N., 
\& Kitayama, T.\ 2013, ApJ, 762, 50

\bibitem[Ciardi 
\& Ferrara(2005)]{Ciardi2005} Ciardi, B., \& Ferrara, A.\ 2005, Space Sci. Rev., 116, 625 

\bibitem[Clarke 
\& Bromm(2003)]{Clarke2003} Clarke, C.~J., \& Bromm, V.\ 2003, MNRAS, 343, 1224

\bibitem[Couchman 
\& Rees(1986)]{Couchman1986} Couchman, H.~M.~P., \& Rees, M.~J.\ 1986, MNRAS, 221, 53

\bibitem[de Souza, Yoshida \& Ioka(2011)]{Souza2011} de Souza, R.~S., Yoshida, N., \& Ioka, K.\ 2011, A\&A, 533, A32 

\bibitem[Dijkstra et al.(2008)]{Dijkstra2008} Dijkstra, M., Haiman, 
Z., Mesinger, A., \& Wyithe, J.~S.~B.\ 2008, MNRAS, 391, 1961

\bibitem[Frebel, Johnson \& Bromm(2009)]{Frebel2009} Frebel, A., Johnson, 
J.~L., \& Bromm, V.\ 2009, MNRAS, 392, L50

\bibitem[Fumagalli, O'Meara \& Prochaska(2011)]{Fumagalli2011} Fumagalli, M., 
O'Meara, J.~M., \& Prochaska, J.~X.\ 2011, Science, 334, 1245

\bibitem[Galli 
\& Palla(1998)]{Galli1998} Galli, D., \& Palla, F.\ 1998, A\&A, 335, 403

\bibitem[Galli 
\& Palla(2002)]{Galli2002} Galli, D., \& Palla, F.\ 2002, Planet. Space Sci., 50, 1197

\bibitem[Glover 
\& Abel(2008)]{Glover2008} Glover, S.~C.~O., \& Abel, T.\ 2008, MNRAS, 388, 1627 

\bibitem[Glover 
\& Brand(2001)]{Glover2001} Glover, S.~C.~O., \& Brand, P.~W.~J.~L.\ 2001, MNRAS, 321, 385

\bibitem[Glover 
\& Brand(2003)]{Glover2003} Glover, S.~C.~O., \& Brand, P.~W.~J.~L.\ 2003, MNRAS, 340, 210

\bibitem[Glover 
\& Jappsen(2007)]{Glover2007} Glover, S.~C.~O., \& Jappsen, A.-K.\ 2007, ApJ, 666, 1

\bibitem[Greif 
\& Bromm(2006)]{Greif2006} Greif, T.~H., \& Bromm, V.\ 2006, MNRAS, 373, 128

\bibitem[Greif et al.(2008)]{Greif2008} Greif, T.~H., Johnson, 
J.~L., Klessen, R.~S., \& Bromm, V.\ 2008, MNRAS, 387, 1021 

\bibitem[Greif et al.(2012)]{Greif2012} Greif, T.~H., Bromm, V., 
Clark, P.~C., et al.\ 2012, MNRAS, 424, 399

\bibitem[Haiman, Rees \& Loeb(1996)]{Haiman1996a} Haiman, Z., Rees, M.~J., 
\& Loeb, A.\ 1996, ApJ, 467, 522 

\bibitem[Haiman, Thoul \& Loeb(1996)]{Haiman1996b} Haiman, Z., Thoul, 
A.~A., \& Loeb, A.\ 1996, ApJ, 464, 523 

\bibitem[Haiman, Rees \& Loeb(1997)]{Haiman1997} Haiman, Z., Rees, M.~J., 
\& Loeb, A.\ 1997, ApJ, 476, 458 

\bibitem[Haiman, Abel \& Rees(2000)]{Haiman2000} Haiman, Z., Abel, T., 
\& Rees, M.~J.\ 2000, ApJ, 534, 11 

\bibitem[Hayakawa, Nishimura \& Takayanagi(1961)]{Hayakawa1961} Hayakawa, S., 
Nishimura, S., \& Takayanagi, T.\ 1961, PASJ, 13, 184

\bibitem[Heger et al.(2003)]{Heger2003} Heger, A., Fryer, C.~L., 
Woosley, S.~E., Langer, N., \& Hartmann, D.~H.\ 2003, ApJ, 591, 288 

\bibitem[Hirano et al.(2014)]{Hirano2014} Hirano, S., Hosokawa, 
T., Yoshida, N., et al.\ 2014, ApJ, 781, 60

\bibitem[Hirasawa, Aizu \& Taketani(1969)]{Hirasawa1969} Hirasawa, T., Aizu, 
K., \& Taketani, M.\ 1969, Progress of Theoretical Physics, 41, 835

\bibitem[Hosokawa et al.(2011)]{Hosokawa2011} Hosokawa, T., Omukai, 
K., Yoshida, N., \& Yorke, H.~W.\ 2011, Science, 334, 1250

\bibitem[Hosokawa et al.(2012)]{Hosokawa2012} Hosokawa, T., Yoshida, 
N., Omukai, K., \& Yorke, H.~W.\ 2012, ApJ, 760, L37

\bibitem[Inayoshi 
\& Omukai(2011)]{Inayoshi2011} Inayoshi, K., \& Omukai, K.\ 2011, MNRAS, 416, 2748 

\bibitem[Inayoshi 
\& Omukai(2012)]{Inayoshi2012} Inayoshi, K., \& Omukai, K.\ 2012, MNRAS, 422, 2539 

\bibitem[Indriolo et al.(2007)]{Indriolo2007} Indriolo, N., Geballe, 
T.~R., Oka, T., \& McCall, B.~J.\ 2007, ApJ, 671, 1736

\bibitem[Inoue, Omukai \& Ciardi(2007)]{Inoue2007} Inoue, S., Omukai, K., 
\& Ciardi, B.\ 2007, MNRAS, 380, 1715 

\bibitem[Inoue 
\& Inutsuka(2008)]{Inoue2008} Inoue, T., \& Inutsuka, S.-i.\ 2008, ApJ, 687, 303

\bibitem[Inoue 
\& Inutsuka(2009)]{Inoue2009} Inoue, T., \& Inutsuka, S.-i.\ 2009, ApJ, 704, 161

\bibitem[Jasche, Ciardi, \& En{\ss}lin(2007)]{Jasche2007} Jasche, J., Ciardi, B., 
\& En{\ss}lin, T.~A.\ 2007, MNRAS, 380, 417

\bibitem[Johnson 
\& Bromm(2006)]{Johnson2006} Johnson, J.~L., \& Bromm, V.\ 2006, MNRAS, 366, 247 

\bibitem[Johnson, Dalla Vecchia \& Khochfar(2013)]{Johnson2013} Johnson, J.~L., Dalla, 
Vecchia,~C., \& Khochfar, S.\ 2013, MNRAS, 428, 1857 

\bibitem[Kang 
\& Shapiro(1992)]{Kang1992} Kang, H., \& Shapiro, P.~R.\ 1992, ApJ, 386, 432

\bibitem[Kitayama et al.(2004)]{Kitayama2004} Kitayama, T., Yoshida, 
N., Susa, H., \& Umemura, M.\ 2004, ApJ, 613, 631

\bibitem[Larson(1969)]{Larson1969} Larson, R.~B.\ 1969, MNRAS, 
145, 271

\bibitem[Longair(2011)]{Longair2011} Longair, M.~S.\ 2011, High 
Energy Astrophysics, by Malcolm S.~Longair, Cambridge, UK: Cambridge 
University Press, 2011

\bibitem[Machacek, Bryan \& Abel(2003)]{Machacek2003} Machacek, M.~E., 
Bryan, G.~L., \& Abel, T.\ 2003, MNRAS, 338, 273 

\bibitem[Machida et al.(2005)]{Machida2005} Machida, M.~N., 
Tomisaka, K., Nakamura, F., \& Fujimoto, M.~Y.\ 2005, ApJ, 622, 39

\bibitem[Mackey, Bromm \& Hernquist(2003)]{Mackey2003} Mackey, J., Bromm, V., 
\& Hernquist, L.\ 2003, ApJ, 586, 1 

\bibitem[McCall et al.(2003)]{McCall2003} McCall, B.~J., 
Huneycutt, A.~J., Saykally, R.~J., et al.\ 2003, Nature, 422, 500

\bibitem[McGreer 
\& Bryan(2008)]{McGreer2008} McGreer, I.~D., \& Bryan, G.~L.\ 2008, ApJ, 685, 8

\bibitem[McKee 
\& Tan(2008)]{McKee2008} McKee, C.~F., \& Tan, J.~C.\ 2008, ApJ, 681, 771

\bibitem[Nagakura 
\& Omukai(2005)]{Nagakura2005} Nagakura, T., \& Omukai, K.\ 2005, MNRAS, 364, 1378

\bibitem[Nagakura, Hosokawa \& Omukai(2009)]{Nagakura2009} Nagakura, T., 
Hosokawa, T., \& Omukai, K.\ 2009, MNRAS, 399, 2183 

\bibitem[Nakamura 
\& Umemura(2002)]{Nakamura2002} Nakamura, F., \& Umemura, M.\ 2002, ApJ, 569, 549 

\bibitem[Nakauchi et al.(2012)]{Nakauchi2012} Nakauchi, D., Suwa, 
Y., Sakamoto, T., Kashiyama, K., \& Nakamura, T.\ 2012, ApJ, 759, 128

\bibitem[Omukai 
\& Nishi(1999)]{Omukai1999} Omukai, K., \& Nishi, R.\ 1999, ApJ, 518, 64 

\bibitem[Omukai(2001)]{Omukai2001} Omukai, K.\ 2001, ApJ, 546, 
635 

\bibitem[Omukai 
\& Palla(2001)]{Omukai_Palla2001} Omukai, K., \& Palla, F.\ 2001, ApJ, 561, L55

\bibitem[Omukai 
\& Palla(2003)]{Omukai_Palla2003} Omukai, K., \& Palla, F.\ 2003, ApJ, 589, 677

\bibitem[Omukai et al.(2005)]{Omukai2005} Omukai, K., Tsuribe, T., 
Schneider, R., \& Ferrara, A.\ 2005, ApJ, 626, 627

\bibitem[Omukai(2007)]{Omukai2007} Omukai, K.\ 2007, PASJ, 59, 
589

\bibitem[O'Shea et al.(2005)]{O'Shea2005} O'Shea, B.~W., Abel, T., 
Whalen, D., \& Norman, M.~L.\ 2005, ApJL, 628, L5

\bibitem[O'Shea et al.(2008)]{O'Shea2008a} O'Shea, B.~W., McKee, 
C.~F., Heger, A., \& Abel, T.\ 2008, First Stars III, 990, 13

\bibitem[O'Shea 
\& Norman(2008)]{O'Shea2008b} O'Shea, B.~W., \& Norman, M.~L.\ 2008, ApJ, 673, 14

\bibitem[Peebles 
\& Dicke(1968)]{Peebles1968} Peebles, P.~J.~E., \& Dicke, R.~H.\ 1968, ApJ, 154, 891

\bibitem[Penston(1969)]{Penston1969} Penston, M.~V.\ 1969, MNRAS, 144, 425

\bibitem[Ritter et al.(2012)]{Ritter2012} Ritter, J.~S., 
Safranek-Shrader, C., Gnat, O., Milosavljevi{\'c}, M., 
\& Bromm, V.\ 2012, ApJ, 761, 56

\bibitem[Rollinde et al.(2008)]{Rollinde2008} Rollinde, E., Maurin, 
D., Vangioni, E., Olive, K.~A., \& Inoue, S.\ 2008, ApJ, 673, 676

\bibitem[Safranek-Shrader, Bromm \& Milosavljevi{\'c}(2010)]{Safranek-Shrader2010} 
Safranek-Shrader, C., Bromm, V., 
\& Milosavljevi{\'c}, M.\ 2010, ApJ, 723, 1568  

\bibitem[Schaerer(2002)]{Schaerer2002} Schaerer, D.\ 2002, A\&A, 382, 28

\bibitem[Shapiro \& Kang(1987)]{Shapiro1987} Shapiro, P.~R., \& Kang, H.\ 1987, ApJ, 318, 32 

\bibitem[Simcoe et al.(2012)]{Simcoe2012} Simcoe, R.~A., Sullivan, 
P.~W., Cooksey, K.~L., et al.\ 2012, Nature, 492, 79

\bibitem[Spitzer 
\& Tomasko(1968)]{Spitzer1968} Spitzer, L., Jr., \& Tomasko, M.~G.\ 1968, ApJ, 152, 971

\bibitem[Spitzer 
\& Scott(1969)]{Spitzer1969} Spitzer, L., Jr., \& Scott, E.~H.\ 1969, ApJ, 158, 161

\bibitem[Stacy 
\& Bromm(2007)]{Stacy2007} Stacy, A., \& Bromm, V.\ 2007, MNRAS, 382, 229 

\bibitem[Stacy, Greif \& Bromm(2012)]{Stacy2012} Stacy, A., Greif, T.~H., 
\& Bromm, V.\ 2012, MNRAS, 42 

\bibitem[Susa(2013)]{Susa2013} Susa, H.\ 2013, ApJ, 773, 185

\bibitem[Tanaka, Moriya \& Yoshida(2013)]{Tanaka2013} Tanaka, M., Moriya, 
T.~J., \& Yoshida, N.\ 2013, MNRAS, 435, 2483 

\bibitem[Tegmark et al.(1997)]{Tegmark1997} Tegmark, M., Silk, J., 
Rees, M.~J., et al.\ 1997, ApJ, 474, 1

\bibitem[Tornatore, Ferrara \& Schneider(2007)]{Tornatore2007} Tornatore, L., 
Ferrara, A., \& Schneider, R.\ 2007, MNRAS, 382, 945 

\bibitem[Trenti 
\& Stiavelli(2009)]{Trenti2009} Trenti, M., \& Stiavelli, M.\ 2009, ApJ, 694, 879 

\bibitem[Tumlinson, Venkatesan \& Shull(2004)]{Tumlinson2004} Tumlinson, J., 
Venkatesan, A., \& Shull, J.~M.\ 2004, ApJ, 612, 602

\bibitem[Uehara 
\& Inutsuka(2000)]{Uehara2000} Uehara, H., \& Inutsuka, S. I.\ 2000, ApJL, 531, L91

\bibitem[Wada 
\& Venkatesan(2003)]{Wada2003} Wada, K., \& Venkatesan, A.\ 2003, ApJ, 591, 38

\bibitem[Webber(1998)]{Webber1998} Webber, W.~R.\ 1998, ApJ, 506, 
329

\bibitem[Whalen, Abel \& Norman(2004)]{Whalen2004} Whalen, D., Abel, T., 
\& Norman, M.~L.\ 2004, ApJ, 610, 14 

\bibitem[Whalen et al.(2013)]{Whalen2013} Whalen, D.~J., Fryer, 
C.~L., Holz, D.~E., et al.\ 2013, ApJL, 762, L6

\bibitem[Wise 
\& Abel(2007)]{Wise2007} Wise, J.~H., \& Abel, T.\ 2007, ApJ, 665, 899

\bibitem[Wise, Turk \& Abel(2008)]{Wise2008} Wise, J.~H., Turk, M.~J., 
\& Abel, T.\ 2008, ApJ, 682, 745 

\bibitem[Wise et al.(2012)]{Wise2012} Wise, J.~H., Turk, M.~J., 
Norman, M.~L., \& Abel, T.\ 2012, ApJ, 745, 50

\bibitem[Wolcott-Green 
\& Haiman(2011)]{Wolcott2011} Wolcott-Green, J., \& Haiman, Z.\ 2011, MNRAS, 412, 2603 

\bibitem[Yamada 
\& Nishi(1998)]{Yamada1998} Yamada, M., \& Nishi, R.\ 1998, ApJ, 505, 148

\bibitem[Yoshida et al.(2006)]{Yoshida2006} Yoshida, N., Omukai, 
K., Hernquist, L., \& Abel, T.\ 2006, ApJ, 652, 6

\bibitem[Yoshida et al.(2007a)]{Yoshida2007a} Yoshida, N., Oh, S.~P., 
Kitayama, T., \& Hernquist, L.\ 2007, ApJ, 663, 687

\bibitem[Yoshida, Omukai \& Hernquist(2007)]{Yoshida2007b} Yoshida, N., Omukai, 
K., \& Hernquist, L.\ 2007, ApJ, 667, L117 

\bibitem[Yoshida, Omukai \& Hernquist(2008)]{Yoshida2008} Yoshida, N., Omukai, 
K., \& Hernquist, L.\ 2008, Science, 321, 669 

\end{thebibliography}
\end{document}